\documentclass[reprint,amsmath,amssymb,apr,aip,onecolumn]{revtex4-2}
\usepackage{graphicx}
\usepackage{graphics}
\usepackage{algorithm,algorithmic}
\usepackage{mathptmx}
\usepackage{times}
\usepackage{amsmath}
\usepackage{amssymb}
\usepackage{dcolumn}
\usepackage{color}
\usepackage{bm}
\usepackage{units}
\usepackage{booktabs}

\newcommand{\GSET}{G\textsubscript{SET}}
\newcommand{\GRESET}{G\textsubscript{RESET}}

\newcommand{\Gmax}{G\textsubscript{max}}
\newcommand{\Wmax}{W\textsubscript{max}}
\newcommand{\What}{\hat{W}}
\newcommand{\yexp}{y\textsubscript{exp}}
\newcommand{\yfp}{y\textsubscript{fp}}
\newcommand{\etot}{\epsilon\textsubscript{total}}
\newcommand{\elin}{\epsilon\textsubscript{linear}}
\newcommand{\eres}{\epsilon\textsubscript{residual}}
\begin{document}
\title{A 64-core mixed-signal in-memory compute chip based on phase-change memory for deep neural network inference}

\author{Manuel Le Gallo}\email{anu@zurich.ibm.com}\affiliation{These authors contributed equally}\affiliation{IBM Research Europe, 8803 R\"{u}schlikon, Switzerland}
\author{Riduan Khaddam-Aljameh}\affiliation{These authors contributed equally}\affiliation{IBM Research Europe, 8803 R\"{u}schlikon, Switzerland}
\author{Milos Stanisavljevic}\affiliation{These authors contributed equally}\affiliation{IBM Research Europe, 8803 R\"{u}schlikon, Switzerland}
\author{Athanasios Vasilopoulos}\affiliation{IBM Research Europe, 8803 R\"{u}schlikon, Switzerland}
\author{Benedikt Kersting}\affiliation{IBM Research Europe, 8803 R\"{u}schlikon, Switzerland}
\author{Martino Dazzi}\affiliation{IBM Research Europe, 8803 R\"{u}schlikon, Switzerland}
\author{Geethan Karunaratne}\affiliation{IBM Research Europe, 8803 R\"{u}schlikon, Switzerland}
\author{Matthias Br\"{a}ndli}\affiliation{IBM Research Europe, 8803 R\"{u}schlikon, Switzerland}
\author{Abhairaj Singh}\affiliation{IBM Research Europe, 8803 R\"{u}schlikon, Switzerland}
\author{Silvia M. M\"{u}ller}\affiliation{IBM Systems and Technology, 71034 B\"{o}blingen, Germany}
\author{Julian B\"{u}chel}\affiliation{IBM Research Europe, 8803 R\"{u}schlikon, Switzerland}
\author{Xavier Timoneda}\affiliation{IBM Research Europe, 8803 R\"{u}schlikon, Switzerland}
\author{Vinay Joshi}\affiliation{IBM Research Europe, 8803 R\"{u}schlikon, Switzerland}
\author{Urs Egger}\affiliation{IBM Research Europe, 8803 R\"{u}schlikon, Switzerland}
\author{Angelo Garofalo}\affiliation{IBM Research Europe, 8803 R\"{u}schlikon, Switzerland}
\author{Anastasios Petropoulos}\affiliation{University of Patras, 26504 Rio Achaia, Greece}
\author{Theodore Antonakopoulos}\affiliation{University of Patras, 26504 Rio Achaia, Greece}
\author{Kevin Brew}\affiliation{IBM Research - Albany, NY 12203, USA}
\author{Samuel Choi}\affiliation{IBM Research - Albany, NY 12203, USA}
\author{Injo Ok}\affiliation{IBM Research - Albany, NY 12203, USA}
\author{Timothy Philip}\affiliation{IBM Research - Albany, NY 12203, USA}
\author{Victor Chan}\affiliation{IBM Research - Albany, NY 12203, USA}
\author{Claire Silvestre}\affiliation{IBM Research - Albany, NY 12203, USA}
\author{Ishtiaq Ahsan}\affiliation{IBM Research - Albany, NY 12203, USA}
\author{Nicole Saulnier}\affiliation{IBM Research - Albany, NY 12203, USA}
\author{Vijay Narayanan}\affiliation{IBM Research - Yorktown Heights, NY 10598, USA}
\author{Pier Andrea Francese}\affiliation{IBM Research Europe, 8803 R\"{u}schlikon, Switzerland}
\author{Evangelos Eleftheriou}\affiliation{IBM Research Europe, 8803 R\"{u}schlikon, Switzerland}
\author{Abu Sebastian}\email{ase@zurich.ibm.com}\affiliation{IBM Research Europe, 8803 R\"{u}schlikon, Switzerland}

\date{\today}
\begin{abstract}
The need to repeatedly shuttle around synaptic weight values from memory to processing units has been a key source of energy inefficiency associated with hardware implementation of artificial neural networks\cite{murmann2021}. Analog in-memory computing (AIMC) with spatially instantiated synaptic weights holds high promise to overcome this challenge, by performing matrix-vector multiplications (MVMs) directly within the network weights stored on a chip to execute an inference workload\cite{shafiee2016,Y2020sebastianNatNano,shimengCAS2021,lanza2022,wang2020,marinella2020}. 
However, to achieve end-to-end improvements in latency and energy consumption, AIMC must be combined with on-chip digital operations and communication to move towards configurations in which a full inference workload is realized entirely on-chip. Moreover, it is highly desirable to achieve high MVM and inference accuracy without application-wise re-tuning of the chip. Here, we present a multi-core AIMC chip designed and fabricated in 14-nm complementary metal-oxide-semiconductor (CMOS) technology with backend-integrated phase-change memory (PCM). The fully-integrated chip features 64 256x256 AIMC cores interconnected via an on-chip communication network. 
It also implements the digital activation functions and processing involved in ResNet convolutional neural networks and long short-term memory (LSTM) networks. 
We demonstrate near software-equivalent inference accuracy with ResNet and LSTM networks while implementing all the computations associated with the weight layers and the activation functions on-chip. The chip can achieve a maximal throughput of \unit[63.1]{TOPS} at an energy efficiency of \unit[9.76]{TOPS W$^{-1}$} for 8-bit input/output matrix-vector multiplications.
\end{abstract}
\maketitle

 Early works on performing neural network inference with AIMC showed promising accuracy results in mixed hardware/software implementations, where functionalities such as digital-to-analog and analog-to-digital conversions, activation functions, and other necessary digital operations were implemented with off-chip software or hardware \cite{Y2016yuIEDM,hu2018,tsai2019,yao2020,Joshi2020}. Since then, various single- or few-core chips using SRAM\cite{biswas2019}, Flash\cite{merrikh2017}, RRAM\cite{cai2019,chen2019,shimeng2020}, PCM\cite{Y2022-khaddam-aljameh-JSSC}, and MRAM\cite{deaville2021} memory technologies have been fabricated with fully-integrated data conversions and sometimes activation functions\cite{cai2019,merrikh2017,Y2022-khaddam-aljameh-JSSC}, demonstrating a competitive energy efficiency of \unit[$ \geq 10$]{TOPS W$^{-1}$} for reduced-precision MVM. In such works, only small networks ($\lesssim 100$k weights) on simple benchmark tasks such as MNIST digit recognition have been implemented to fit entirely on the core\cite{cai2019,chen2019,merrikh2017,Y2022-khaddam-aljameh-JSSC,shimeng2020}, whereas larger networks have been emulated by programming each layer one at a time onto the same core\cite{deaville2021,biswas2019}. Recently, multi-core chips supporting larger networks ($> 1$M weights) have demonstrated promising inference accuracy results on more difficult benchmarks, such as image recognition on the CIFAR dataset, and high MVM energy efficiency  \cite{Khwa2022,wan2022compute,hung2021four,mythic2022,Jia2022,ares2021}. 

Despite those advances, significant challenges remain to build a complete AIMC accelerator to demonstrate end-to-end inference tasks with accuracy and efficiency that outperform the already existing digital accelerators. Because of the rather large size of the multi-bit SRAM plus switched-capacitor unit-cells, the existing SRAM-based implementations need off-chip weight buffers to hold network weights, which are then partially transferred to the AIMC cores because an entire network cannot fit fully on-chip \cite{Jia2022}. The current implementations are not dense enough to hold $>10$MB of weight data at reasonable chip area, limiting their efficiency gains over digital approaches to rather small networks that fit entirely on-chip \cite{murmann2021}. Another limitation is the volatility of the SRAM cells, which prevent the on-chip weights to be retained throughout power cycling of the accelerator. 

The potential to achieve smaller unit-cell sizes with resistive memory technologies could make it possible to hold such large networks entirely on-chip in a non-volatile way and eliminate the overhead of off-chip weight data access. In this context, an AIMC accelerator based on such devices becomes a natural choice. Both digital\cite{Khwa2022,hung2021four} (1 device corresponds to 1 weight bit) and analog (1 device encodes an entire weight)\cite{wan2022compute,ares2021} storage of weights in those devices have been explored in the recent AIMC chips. Digital storage offers high accuracy, but the throughput is limited due to the small number of rows that can be operated at a time (sometimes as low as 4)\cite{hung2021four} without degrading the signal quality, in order to prevent large errors when processing the most significant bits. This approach also limits the achievable weight density because several devices are needed to encode a weight. Analog weight storage offers high weight density and the ability to operate more rows simultaneously (demonstrated up to 512)\cite{ares2021}. However, this approach suffers from accuracy degradation because of the noisy analog weights and higher latency due to the need of slow high resolution analog-to-digital converters (ADCs) that often have to be multiplexed across columns. Moreover, extensive network-specific chip tuning, such as per-network recalibration of the chip to mitigate non-idealities of the devices and peripheral circuitry, and chip-in-the-loop finetuning of weights through backpropagation, has so far been necessary to achieve acceptable inference accuracy with this approach\cite{wan2022compute,yao2020}. Finally, at the system level, none of the current AIMC chips based on  resistive memory support all the computations involved in processing convolutional and LSTM layers. 

To address those challenges, we present the IBM HERMES Project Chip: a 64-core AIMC chip based on back-end integrated PCM in a 14-nm CMOS process. The chip delivers simultaneously four key advances over the previous implementations discussed above. (i) High weight capacity is enabled through the analog storage capability of PCM, using only 4 PCM devices to encode a weight. (ii) Low MVM latency, high parallelism, and high throughput are achieved by using compact current-controlled oscillator ADCs placed on each unit-cell row to perform fully-parallel MVMs. (iii) High MVM and inference accuracy are achieved without application-wise chip re-tuning, thanks to proper on-chip ADC calibration circuits that remove unwanted nonlinear response. (iv) Finally, all non-MVM computations involved in processing convolutional and LSTM layers are supported using dedicated high-precision digital units. We report measured inference results at near software-equivalent accuracy on CIFAR-10 image recognition, Penn Treebank (PTB) character prediction, and image caption generation tasks with all convolutional and LSTM layer computations performed on-chip. The measured MVM throughput per area of the chip is $>15\times$ higher than comparable multi-core resistive memory AIMC chips, while achieving the highest accuracy on the CIFAR-10 benchmark. 

\section{Chip architecture}
\begin{figure*}[t!]
\centering
\begin{tabular}{c}
\includegraphics[width = 0.99\columnwidth]{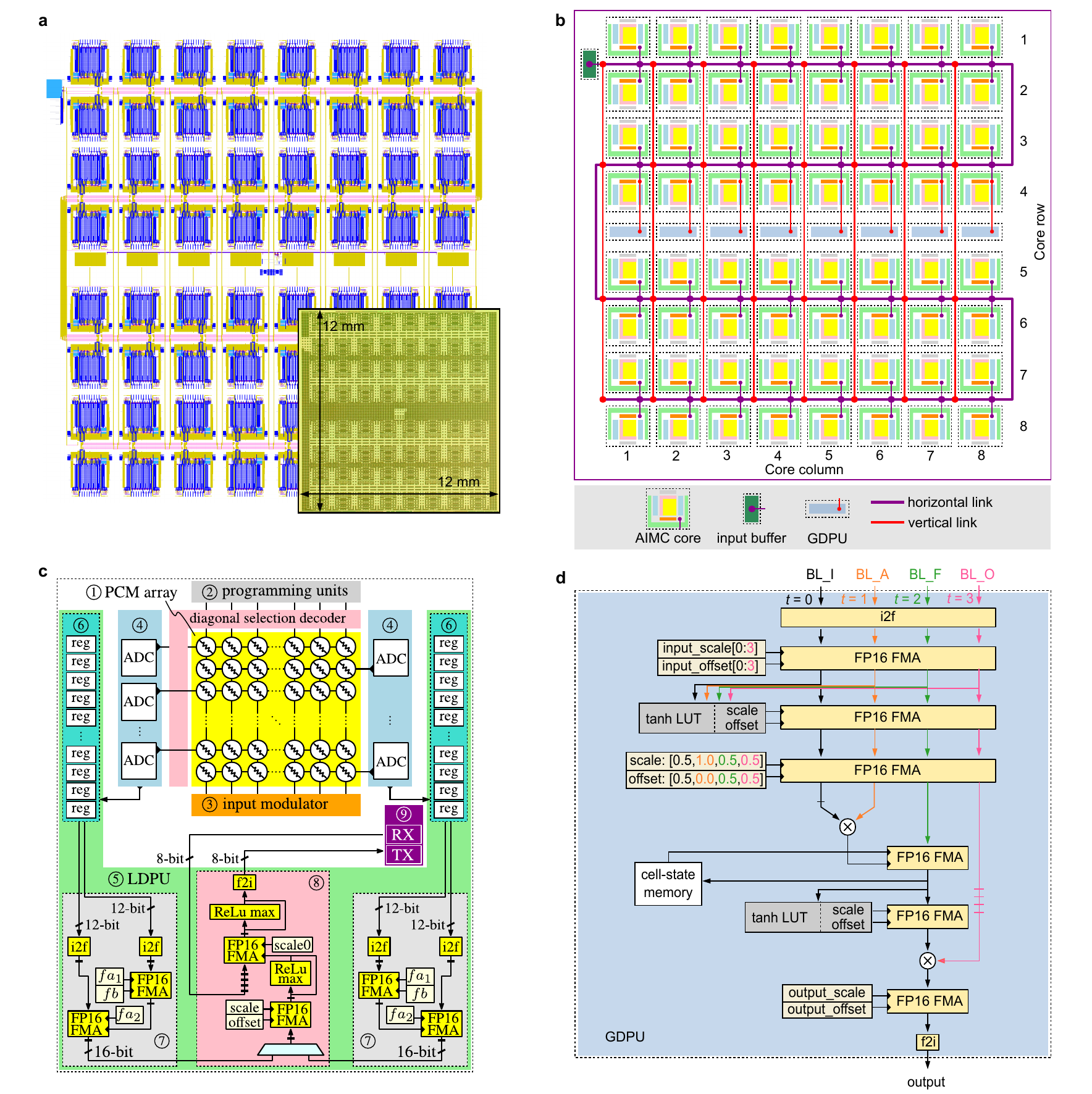}
\end{tabular}
\caption{\textbf{IBM HERMES Project Chip overview.} \textbf{a}, Electronic design automation snapshot and inset showing a micrograph of the chip. Therein, the outline of the 64 cores can be recognized as well as the array of 5,616 pads. \textbf{b}, Schematic overview of the different components on the multi-core chip. \textbf{c}, Schematic overview of a single PCM-based in-memory compute core. (1) PCM crossbar array, (2) current DAC-based programming unit, (3) PWM-based input modulator, (4) left and right ADC arrays, (5) local digital processing unit (LDPU), (6) left and right ADC register arrays, (7) left and right ADC convert and scale blocks, (8) activation function block, (9) link controller. \textbf{d}, Block diagram of a global digital processing unit (GDPU) used for LSTM-related data processing. Inputs to and outputs from the GDPU slice are in an 8-bit signed integer format (\texttt{INT8}). By using custom conversion blocks marked by \texttt{i2f} and \texttt{f2i}, \texttt{INT8} values can be converted into \texttt{FP16} and vice versa. Additionally, conversions at input/output can encompass a per-gate/per-output scale and bias operation using the FMA units. The inputs from the I, A, F, and O BLs are time multiplexed, and a single block is used to compute the gates' activation vectors. The sigmoid activation function for the I, F and O gates is computed by scaling and offsetting the output of the hyperbolic tangent function with the third (from the top) FMA unit, according to the identity $\mathrm{sigmoid}(x)=1/2 + 1/2  \cdot \tanh(x/2)$. 
} \label{fig:1}
\end{figure*}

Fig.~\ref{fig:1}a shows a snapshot of the chip inside the electronic design automation (EDA) tool.  
The chip has a square dimension of $\unit[12]{mm}\times\unit[12]{mm}$. 
Its 64 cores, each of size $\unit[1.2]{mm}\times\unit[1.16]{mm}$, are placed in eight rows and eight columns.
This subdivided structure is detailed in the schematic illustration in Fig.~\ref{fig:1}b.
Each core contains a PCM crossbar array capable of storing a $256 \times 256$ weight matrix and performing an analog MVM using the input activations provided from outside of the core. Hence, up to 4,194,304 weights can be stored on-chip for performing in-memory MVMs. 

At the center of the core grid, between rows four and five, a row of eight global digital processing units~(GDPUs) is placed. 
They provide the additional digital post-processing capabilities needed when running LSTM networks.
The high-frequency clock signals for the digital circuitry in the cores and GDPUs are received at the center of the chip and are distributed in a balanced tree across the chip.

Finally, the core outputs and GDPU inputs are interconnected by a grid of horizontal and vertical digital communication links (see Fig.~\ref{fig:1}b), yielding a DNN acceleration architecture where weight elements are stationary, and only the activation vectors are transmitted. There are 418 physical communication links in total, which together implement a 5 Parallel Prism communication fabric topology\cite{dazzi2021} (see Extended Data Fig.~\ref{extfig_link} for a map of all possible link connections).
A summary of the chip specifications can be found in Extended Data Table \ref{exttab:1}.

\section{The computational memory core}
Fig.~\ref{fig:1}c depicts the architecture of a single core and its various components.
The design is based on the fully-integrated standalone PCM-based single-core chip presented in Ref.~\onlinecite{Y2022-khaddam-aljameh-JSSC}, with the addition of the link controller for transmitting and receiving data across different cores. At the center of the core, there is a $256 \times 256$-sized crossbar array of PCM-based unit-cells.
As shown in Extended Data Fig.~\ref{extfig_core}a, four PCM devices are used per unit-cell: two each for each polarity, such that positive and negative weights can be represented~\cite{Boybat2018,Y2021-khaddam-aljameh-TCASII}. The phase-configuration of the phase-change material within the mushroom-type PCM devices determine their conductance (see Extended Data Fig.~\ref{extfig_device}).
The devices in a conductance pair are weighted with equal significance. 

To program the individual conductances, a diagonal decoding scheme~\cite{Y2021-khaddam-aljameh-TCASII} is employed to select $256$ different devices, one per row and column. 
The whole procedure starts with the dedicated per-core programming finite-state machine (FSM) instructing the diagonal selection decoder to enable one diagonal of cells that contains the devices that are to be programmed (see Extended Data Fig.~\ref{extfig_core}b).
Subsequently, these selected devices can be programmed by the current-steering digital-to-analog converter~(DAC) based programming units located on top of the PCM array. 
The DAC units are capable of producing all required programming pulse shapes for the PCM devices based on the instruction provided by the FSM. 
This includes high current \textsc{reset} pulses, triangular-shaped \textsc{set} pulses, and programming pulses with variable current amplitude and time duration.

Once all programming operations are concluded, the core can be used to perform analog MVMs. To this end, the input modulator applies pulse-width modulated~(PWM) read voltage pulses to the PCM array, which are digitized by an array of 256 time-based current-ADCs.
These ADCs flank the PCM-array from the left and right side, such that each unit-cell row is associated with a dedicated ADC.
This is achieved by matching the ADC height to that of two unit-cells, which results in a skewed aspect ratio, as is shown in the layout in Extended Data Fig.~\ref{extfig_core}c. By choosing a current-based ADC topology, area-intensive and noise-prone current-to-voltage conversion steps can be avoided. Furthermore, by opting for a time-based architecture, a highly digital design is obtained in which the output precision can be adjusted by changing the conversion time.

To counteract manufacturing mismatch and other non-idealities, the ADC includes an internal resistor ladder-based voltage DAC that allows correcting the read-voltage offset. Furthermore, the internal current mirror, which drives the attached current-controlled oscillator (CCO) unit, contains trimming registers that allow matching the gain between the different ADCs per core and compensating nonlinearity in their transfer function\cite{Y2022-khaddam-aljameh-JSSC}.
Thus, after performing an initial calibration procedure, the configuration parameters are stored on-chip such that the ADCs exhibit tightly distributed and linear transfer curves. Note that this calibration is independent of the deployed application and does not have to be redone for different weight sets or networks.


Extended Data Fig.~\ref{extfig_read}a illustrates one row of the computational memory during an analog MVM.
The per-row ADC regulates the bit line (BL) to a fixed common-mode voltage, such that the top electrodes of all PCM devices in the 256 unit-cells are on the same potential.
Their respective source line (SL) potentials, however, are connected to one of three potentials based on the respective input value, i.e., the positive read voltage $V^+$ to subtract current, the negative read voltage $V^{-}$ when adding current, or high impedance when no current from the cell is to be added.
Thus, all combinations of products between a positive or negative weight and a positive or negative input can be performed in the analog domain in a single modulation, as shown in Extended Data Fig.~\ref{extfig_read}b.
In addition, a 4-phase mode is supported that is shown in Extended Data Fig.~\ref{extfig_read}c. 
Therein, the four combination of $\pm$weights and $\pm$inputs are performed in four separate operations while only using the negative read voltage $V^{-}$.
As a result, a higher accuracy of the measured current is achieved, as all PCM devices are read in the same polarity by the current-based ADC. Therefore, all experiments in this work were performed using the higher precision 4-phase mode. Compensation schemes to mitigate the impact of the current-voltage polarity dependence of PCM\cite{ghazi2022} to enable high precision in the 1-phase mode will be the subject of future investigations. 

The ADC delivers the final results as two 12-bit unsigned integers, one for the positive and one for the negative current.
By using tri-state buffers on a 24-bit wide bus, the results are transmitted from each ADC to the local digital processing unit~(LDPU) for post-processing.
In the first step, the results are transferred to a register array, such that the operation becomes fully pipelined.
Consequently, the initial latency of the LDPU does not affect its throughput (one data point per clock cycle).
Next, the results are transferred one by one to the ADC-convert-and-scale block, which first converts the 12-bit integers to half-precision floating-point format~(\texttt{FP16}) and then employs two fused multiply-add~(FMA) units to remove gain and offset errors.

Two ADC-convert-and-scale blocks produce outputs every second clock cycle and feed them one after the other to the activation function block through a multiplexer unit. 
There, affine scaling per channel, as well as an optional ReLU operation, are applied. 
The intermediate value is then combined with an 8-bit signed integer~(\texttt{INT8}) input coming from the neighboring cores, or from off-chip, through the link controller receiver~(RX). This value is first converted into \texttt{FP16} format and then scaled to the optimal range. 
Finally, the output undergoes another optional ReLU operation before conversion into \texttt{INT8}, ready to be shipped by the transmitter~(TX) of the link controller. 
The existence of both ReLU operations allows the possibility to apply the activation either before or after adding the RX link input.
The activation block is fully pipelined so that its latency does not affect the LDPU throughput.

The link controller implements a state machine that enables transmission and reception of data within at most six neighboring cores (receiving and transmitting cores are not necessarily the same, see Extended Data Fig.~\ref{extfig_link}). The core-to-core input-output links are implemented as eight parallel channels, each transferring one bit every clock cycle. The selection of the data source for the TX and of the data destination for the RX is implemented through a routing table, configured before runtime. Specifically, the transmitter-side link controller prepends a preamble to the payload being sent on the links. The preamble contains information to uniquely identify a routing path for a particular type of payload. If the receiving core is enabled to receive data from the transmitting core in the routing table, the link controller samples the incoming data. Furthermore, the link controller in the transmitting and receiving cores can select a portion of the payload according to another set of routing registers. In the current design, the links can transfer data across the LDPUs of multiple cores to realize fully on-chip intra-layer partial sum accumulation, and transfer data from the LDPU to the input of a GDPU.

\section{The global digital processing unit}
The GPDU is responsible for the digital processing required in LSTM networks. 
The chip contains eight GDPU slices in total, each connected to the fourth row core above a GDPU slice. 
The cores in a column above a GDPU slice can aggregate their outputs in a cascaded fashion in the LDPUs of those cores. 
The final aggregated output produced by the LDPU of the fourth row core is then sent to its corresponding GDPU slice. 
Each slice takes up to 256 inputs serially (one in each clock cycle), out of which 64 inputs of input gate's (I), cell input (A), forget gate's (F), and output gate's (O) pre-activation vector. 
Those pre-activation vector inputs arrive in an alternating fashion from four neighbouring BLs.
By mapping columns of the I, A, F, and O weight matrices in an interleaved fashion, a single element processor per GDPU slice (instead of 64) can be used to process the 256 inputs serially, which significantly reduces the number of resources needed in the GDPU.
Each GDPU slice produces one output every four clock cycles for up to 64 outputs.
It also contains cell state memory registers for up to 64 elements.

Fig.~\ref{fig:1}d depicts one GDPU slice architecture in detail. 
All computations inside the GDPU slice are performed in the \texttt{FP16} format with inputs and outputs in \texttt{INT8} format.
Hyperbolic tangent activation functions are realized using a lookup table (LUT) method. 
First, a bank of 17 comparators is used to find the corresponding bin (out of 18) in which an input is contained. 
Then, the output value is interpolated in a fused-multiply-add (FMA) unit using the slope and offset read from the LUT for the determined bin. 
The output activations of the four gates are further processed by two multiply blocks, three FMA blocks, and one more hyperbolic tangent LUT block to compute the cell-state and the hidden state output of the LSTM unit.
%
%

\section{MVM accuracy}

\begin{figure*}[t!]
\centering
\begin{tabular}{c}
\includegraphics[width = 0.99\columnwidth]{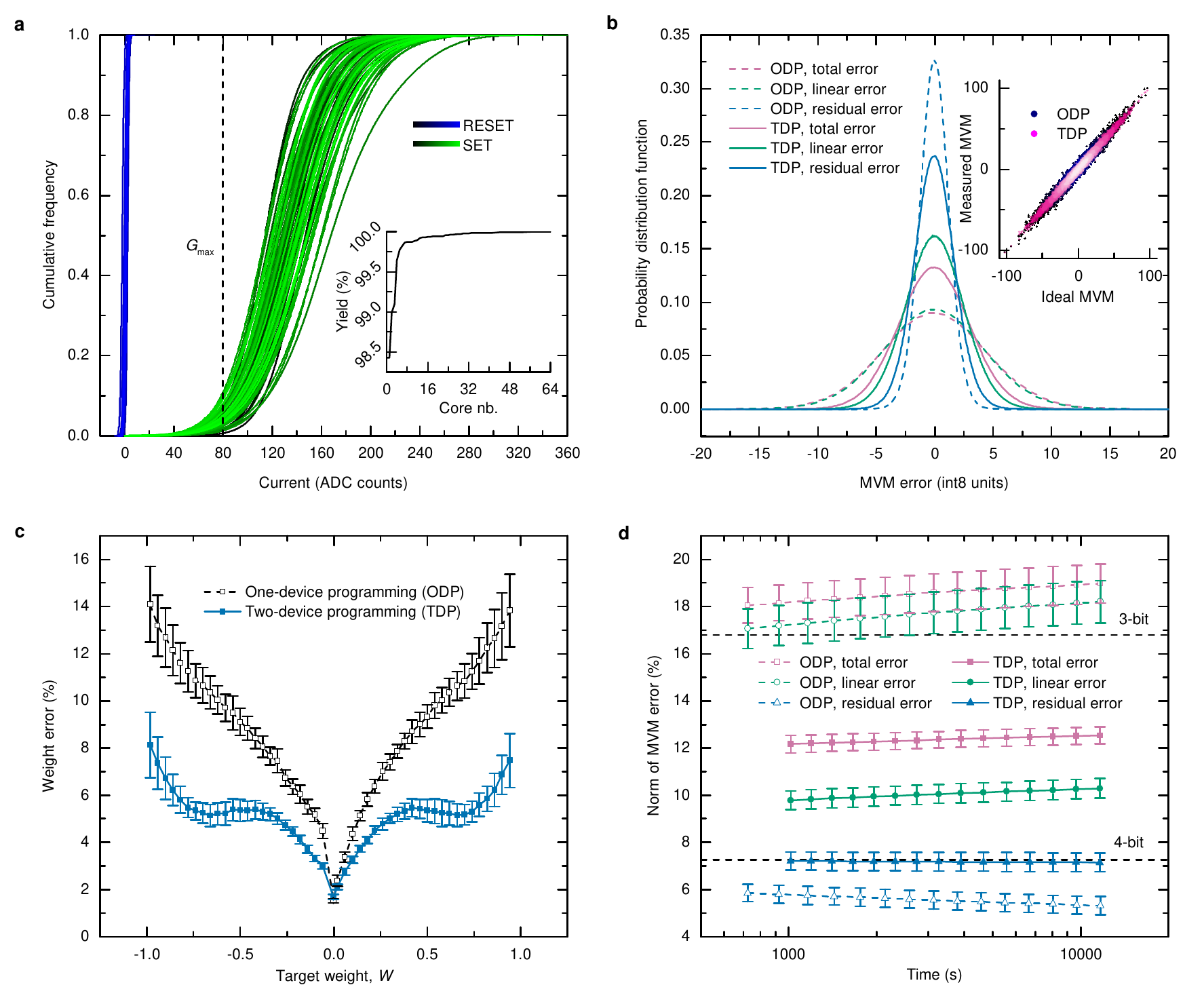}
\end{tabular}
\caption{\textbf{MVM characterization}. \textbf{a}, Unit-cell SET and RESET distributions of the 64 cores. Shades of blue and green denote different cores. The dashed line represents the one-device programming (ODP) $\Gmax$, defined as the tenth percentile of the core with the least conductive SET states. The inset shows the unit-cell yield of the 64 cores. The yield condition is that the unit-cell can be programmed to $|\GRESET|<5$ and $\GSET>50$. \textbf{b}, Error distributions of $\etot$, $\elin$ and $\eres$ for the 64 cores in LDPU (int8) units. A uniformly distributed weight matrix with 30\% sparsity is programmed on each core, and 2,048 input vectors uniformly distributed with 30\% sparsity are then sent to each core to perform the MVMs. The reduction of $\etot$ achieved with two-device programming (TDP) can be attributed to a reduction of $\elin$. The distributions are computed over all error vector elements of 2,048 MVMs performed with all 64 cores. The inset shows the measured MVM results of one core for ODP and TDP against the ideal MVMs computed in software. \textbf{c}, Weight error $\mathrm{std}(W-\hat{W})/\Wmax$, where $\mathrm{std}(W)$ is the standard deviation computed over all elements of $W$, as a function of target weight for ODP and TDP. The error bars represent one standard deviation over the 64 cores. \textbf{d}, 2-norm of $\etot$, $\elin$ and $\eres$, normalized by $|| \yfp ||_2$, as a function of time for ODP and TDP. Due to temporal conductance drift of PCM devices, $\etot$ and $\elin$ increase gradually. The dashed lines represent the error achieved by a digital engine with 8-bit input/output precision and $N$-bit weight precision. The error bars represent one standard deviation over the 64 cores.}   
\label{fig:2}
\end{figure*}

We characterized all 64 cores of a single chip to assess the PCM yield and how accurately MVMs can be performed with them. 
Fig.~\ref{fig:2}a shows the SET and RESET distributions of the unit-cells of all 64 cores. The unit-cell comprises four PCM devices (two per polarity). Therefore, we define a unit-cell in the RESET state ($\GRESET$) when all PCM devices are programmed to RESET, and a unit-cell in the SET state ($\GSET$) when one PCM device is programmed to SET and the others to RESET. On 63 out of the 64 cores of this chip, more than 99\% of the unit cells can be programmed to $|\GRESET|$ smaller than 5 ADC counts and $\GSET$ larger than 50 ADC counts. The yield of the remaining outlier core is 98.4\%. The high yield achieved on all cores implies that it is possible to program most of the unit-cells to a given conductance value via iterative programming, as long as it is within the range of their $\GRESET$ and $\GSET$.

We studied two programming algorithms using either strictly one device (ODP) or up to two devices (TDP) to map a weight to a unit-cell conductance. A weight is written by programming the unit-cell devices that correspond to the weight polarity. Devices of the opposite polarity are RESET to near zero conductance. For a given unit-cell at row $m$ and column $n$ in a core, the corresponding weight $W_\mathrm{mn}$ is mapped to a target conductance value $G_\mathrm{mn}$ according to
\begin{equation}
G_\mathrm{mn} = W_\mathrm{mn} \cdot \frac{\Gmax}{\Wmax}, 
\label{eq:weight_to_G}
\end{equation}
where $\Wmax$ is the maximum absolute weight value to be programmed in the core, and $\Gmax$ is the maximum reliably programmable unit-cell conductance. For ODP, $\Gmax$ should ideally not be larger than the least conductive SET device of the core. We define $\Gmax = 80$ ADC counts based on SET distributions of Fig.~\ref{fig:2}a, as the tenth percentile of the core with the least conductive SET states. For TDP, $\Gmax$ should be set according to the least conductive unit-cell when both devices of positive or negative polarity are in SET state. Here, we choose $\Gmax = 160$ ADC counts. 

We use a closed-loop iterative programming scheme in which the programming current is updated proportional to the difference between the read conductance of unit-cell $(m,n)$ and $G_\mathrm{mn}$ \cite{Y2011papandreouISCAS}. In each iteration, the devices are read individually, and a programming pulse is applied to the unit-cells that did not yet converge within a pre-defined margin. For ODP, one out of the two unit-cell devices corresponding to the weight polarity receives the iterative programming pulses, and the other is RESET. For TDP, if $G_\mathrm{mn}$ exceeds $\GSET$ of both devices, one is programmed to the SET state and the other receives the iterative programming pulses. Otherwise, the device with the highest $\GSET$ receives the iterative programming pulses and the other is RESET (see Methods and Extended Data Fig.~\ref{extfig_prog}). Performing the device selection this way ensures that a maximal amount of devices are at RESET or SET, which are the least noisy states. Because of this, reduced programming errors are achieved compared with prior multi-device programming approaches that attempt to program each unit-cell device via iterative programming up to the ODP $\Gmax$ \cite{legallo2022} (see Extended Data Fig.~\ref{extfig_prog}c).

To quantify the MVM accuracy achieved with ODP and TDP on the 64 cores, a randomly distributed weight matrix $W$ is programmed on each core, and 2,048 randomly distributed input vectors $x$ are then sent to each core to perform the MVMs. Based on the measured MVM results $\yexp$, the actually programmed weights can be estimated as $\hat{W} = \mathrm{argmin}_{\hat{W}}||\yexp - \hat{W} x||_2$. We then characterize the MVM error using the three following metrics: 
\begin{eqnarray}
\etot &=& \yexp - \yfp  \label{eq:etot}\\
\elin &=& \What x - \yfp \label{eq:elin}\\
\eres &=& \yexp - \What x \label{eq:eres},
\end{eqnarray}
where $\yfp = Wx$ denotes the MVM result computed in double precision floating-point. $\etot$ is the total MVM error, which we decompose into a linear contribution $\elin$ and a residual contribution $\eres$. In a first approximation, $\elin$ defines the error resulting from a wrong programming of the weights. $\eres$ results from any residual error of the chip that cannot be cast as a weight error, such as PWM and ADC nonlinearities, instantaneous read noise, leakage current and IR drop. 

The error histograms of the three metrics over all 64 cores are plotted in Fig.~\ref{fig:2}b for ODP and TDP. For ODP, $\etot$ is largely defined by $\elin$, and $\eres$ is comparably negligible. When using TDP, $\elin$ is significantly reduced compared with ODP, showing that the weight programming error is effectively lowered when using more devices per weight. In this case, the impact of $\eres$ gets noticeable in the total error. We ascribe the wider spread of $\eres$ for TDP to the fact that a larger range of the ADC transfer curve and particularly the more nonlinear high-current-range is sampled. Moreover, higher leakage from the PCM array than ODP is expected because more devices are programmed to non-RESET states, which also increases $\eres$. Fig.~\ref{fig:2}c shows the weight error plotted as a function of target weight, confirming that TDP allows to reduce it for all nonzero weight values. We attribute this reduction to a higher signal-to-noise ratio achieved when using two devices instead of one \cite{legallo2022}. 
 
For resistive memory technologies, it also critical to study the degradation of the MVM accuracy over time. In PCM, the conductance values drift with time $t$ according to $G(t) = G(t_0) \left( t / t_0 \right)^{-\nu}$, where $G(t_0)$ is the conductance at $t_\mathrm{0}$ and $\nu$ denotes the drift exponent\cite{Ielmini2009,Y2018legalloAEM}. Drift can partially be mitigated by a global drift compensation procedure, i.e., an appropriate per-core rescaling the MVM result \cite{legalloTED2018,legallo2022} . However, since each programmed weight drifts with a slightly different rate ($\nu$ depends on the conductance state and varies from device to device), $\elin$ and thus also $\etot$ increase over time as shown in Fig.~\ref{fig:2}d. The subtle reduction of $\eres$ can be ascribed to the sampling of a gradually smaller range of the ADC transfer curve. Compared with an equivalent digital engine having 8-bit input/output precision and $N$-bit weight precision, the precision achieved with ODP is close to 3-bit weights. With TDP, a precision between 3-bit and 4-bit weights is achieved.

\section{Deep neural network inference demonstrations}

Having established the MVM accuracy of the chip, we now assess the inference accuracy when running different neural networks on it. We implemented three networks that cover all usable features of the chip: a ResNet-9 convolutional neural network (CNN) for CIFAR-10\cite{cifar} image classification, an LSTM network for predicting characters of the PTB dataset\cite{ptb}, and an LSTM network for generating captions of images in the Flickr8k dataset\cite{flickr} (see Methods). Prior to being deployed on the chip, the networks are trained in a hardware-aware manner by injecting noise on the synaptic weights to improve their resilience to hardware nonidealities \cite{Joshi2020}, using the publicly available IBM Analog Hardware Acceleration Kit \cite{aihwkit} (see Methods for details about the training procedure). The trained weights are then mapped to conductance values to be programmed on the PCM unit-cells using Eqn.~\eqref{eq:weight_to_G}. $\Gmax$ is determined on a per-core basis while constraining the BL current to not exceed the maximum supported by the ADC (see Methods). For the three networks, all operations involved in the processing of weight layers, including the MVMs, activation functions, batch normalization, LSTM cell-state computation, data aggregation when a layer is split onto multiple cores or for residual connections, bias addition, and all the necessary intra- and inter-layer affine scaling, are implemented on-chip. Apart from the max-pooling in ResNet-9 and the character/image/word embedding in the LSTM networks, which are not supported by the chip, no other computation is performed in software in the network executions. Only the communication of activation vectors between the output of a layer and the input of the next layer is done by off-chip data transport through the field-programmable gate array on the test board due to limitations in the communication fabric, and will be implemented on-chip in future designs.

\begin{figure*}[t!]
\centering
\begin{tabular}{c}
\includegraphics[width = 0.99\columnwidth]{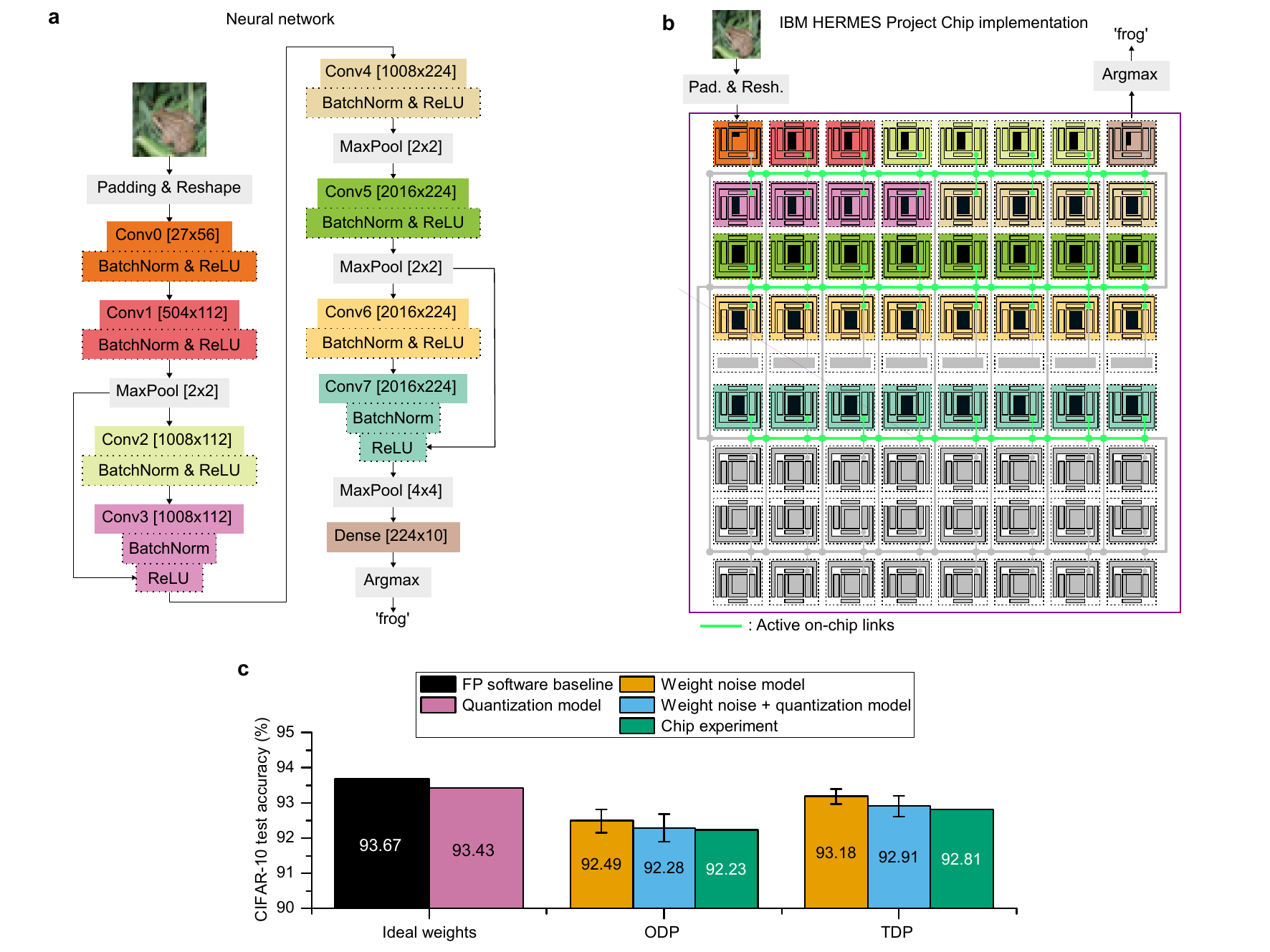}
\end{tabular}
\caption{\textbf{ResNet-9 on CIFAR-10 measurement results. } \textbf{a}, Network architecture. ResNet-9 comprises 8 convolution layers and 1 dense classification layer. Each convolution layer is followed by batch normalization and ReLU activation. Two residual connections from Conv1 to Conv3 and Conv5 to conv7, are present. Four max-pooling layers, implemented off-chip, are used to reduce the size of the input volume along network depth. \textbf{b}, Mapping of ResNet-9 onto the chip. Weights of all layers are programmed onto 40 cores. Conv0-4 and the dense layer are implemented on the first two rows of cores, whereas Conv5-7 each span an entire row. On-chip links connect the LDPUs of the cores within a layer to realize on-chip data aggregation. Batch normalization, ReLU, residual and bias additions are implemented in the LDPUs. \textbf{c}, Measured test accuracy results on CIFAR-10 benchmark for ODP and TDP compared with software baseline and simulation results that include the hardware-measured weight noise and quantization from the PWM and LDPU. The error bars represent one standard deviation over 10 inference runs. } \label{fig:3}
\end{figure*}

The custom ResNet-9 network (1,866,536 synaptic weights and 10 biases) and its corresponding implementation on the chip are shown in Fig.~\ref{fig:3}a and Fig.~\ref{fig:3}b, respectively. On-chip links are configured to aggregate MVM results in the LDPUs of the cores implementing layers that cannot fit onto a single one (all layers except the first and the last). The batch normalization and ReLU activation for all convolution layers are performed in the LDPU of the core that receives the aggregated data. The residual connections are implemented by sending the residual data to the LDPU of the receiving core, and performing the addition inside the LDPU. For the last dense layer, the bias is programmed into the affine scale registers of the LDPU so that it gets added to the MVM result. The CIFAR-10 test accuracy results for ODP and TDP weight programming are shown in Fig.~\ref{fig:3}c. We achieve a hardware accuracy of $92.81 \%$ with TDP, which is less than $1 \%$ below the software baseline of $93.67 \%$, whereas ODP achieves $92.23 \%$. We also compare the experimental results with simulations that include the per-core weight error for this network extracted as in Fig.~\ref{fig:2}c, and the 8-bit quantization from the PWM and LDPUs (see Methods for details on the simulation model). The biggest accuracy drop from the software baseline comes from the weight error, and the digital quantization adds another $0.2-0.3 \%$ drop on top of it (see Fig.~\ref{fig:3}c). The accuracy obtained by simulating only those two effects nearly matches the experimental accuracy, which indicates that any additional nonlinearity from the peripheral circuitry of the chip does not have a significant impact compared with the weight and quantization errors. 

\begin{figure*}[t!]
\centering
\begin{tabular}{c}
\includegraphics[width = 0.99\columnwidth]{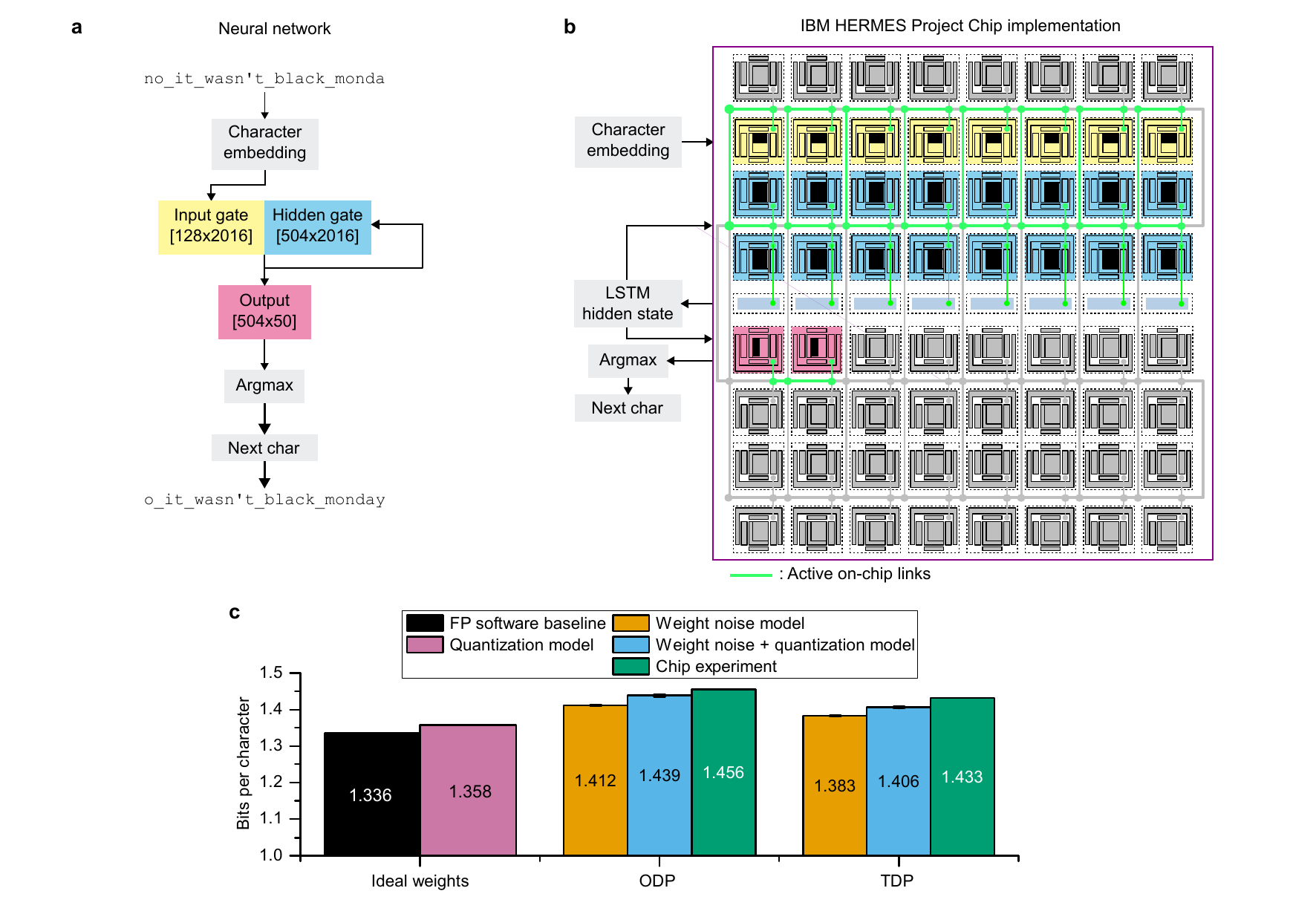}
\end{tabular}
\caption{\textbf{LSTM for character prediction measurement results. } \textbf{a}, Network architecture. The LSTM unit comprises an input gate that receives the embedded input characters and a hidden gate whose input is the output hidden state vector of the LSTM unit. The output layer is a dense classification layer whose output is the next predicted character. \textbf{b}, Mapping of the LSTM network onto the chip. Weights of all layers are programmed onto 26 cores of the chip. The rows 2, 3 and 4 encode the weights LSTM unit and vertical on-chip links connect their LDPUs to realize on-chip data aggregation. The LDPUs of the cores in row 4 are connected to their respective GDPU slice via on-chip links. \textbf{c}, Measured BPC results for ODP and TDP compared with software baseline and simulation results that include the hardware-measured weight noise and quantization from the PWM, LDPU, and GDPU. The error bars represent one standard deviation over 10 inference runs. } \label{fig:4}
\end{figure*}

The LSTM network for PTB character prediction (1,299,312 weights and 2,066 biases) is shown in Fig.~\ref{fig:4}a, and the corresponding implementation on the chip in Fig.~\ref{fig:4}b. Each character is first embedded into a 128-long random orthogonalized vector which is then input to the on-chip input gate. Then, MVMs from the input and hidden gates are summed vertically on-chip by the LDPUs of the 8 cores in rows 3 and 4, using vertical on-chip links to transmit the data. The data produced by the LDPUs of the 8 cores of row 4 are then sent to the 8 corresponding GDPU slices using on-chip links. The GDPU then performs the sigmoid and tanh activations, the cell-state computation, as well as the necessary input/output data scaling. Finally, the output of the GDPU is sent to the dense layer which outputs the next predicted character in the sequence, as well as to the hidden gate as input for the character of the next timestep. The performance of the character prediction task is evaluated in bits per character (BPC), which is a measure of how well the model is able to predict samples from the true underlying probability distribution of the dataset (a lower BPC corresponds to a better model). The BPC results for ODP and TDP weight programming are shown in Fig.~\ref{fig:4}c as well as the corresponding simulation results. With TDP on hardware, we achieve a BPC less than 0.1 higher than the software baseline of 1.336 (adequate baseline on this benchmark for a network of this size \cite{NIPS2017_PTB}), whereas the ODP BPC is only $\sim 0.02$ higher than TDP. Unlike for ResNet-9, the high per-core utilization of the LSTM network leads to a high enough current at the ADC input with ODP to achieve a good signal-to-noise ratio, therefore using TDP only marginally improves the BPC. Here again, the simulated accuracy with weight noise and quantization nearly matches the experimental accuracy.

\begin{figure*}[t!]
\centering
\begin{tabular}{c}
\includegraphics[width = 0.99\columnwidth]{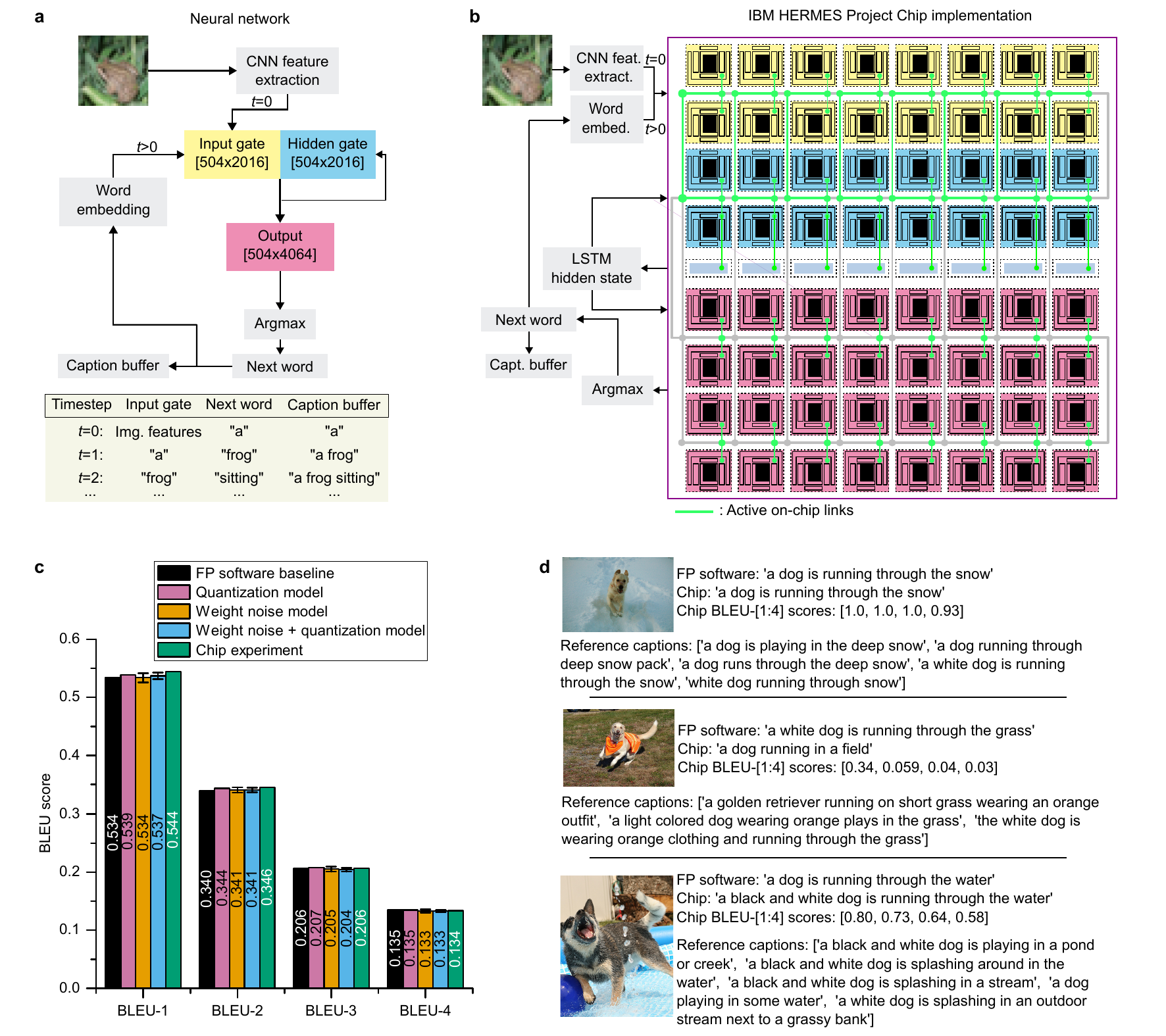}
\end{tabular}
\caption{\textbf{LSTM for image caption generation measurement results. } \textbf{a}, Network architecture. The features of the input image extracted by a CNN (off-chip) are fed to the input gate of the LSTM unit. The output hidden state vector of the LTSM unit is sent to the output dense layer that generates the first word of the caption, which is stored in a buffer. This word is then embedded and fed back to the input gate of the LSTM unit to generate the next word. \textbf{b}, Mapping of the network onto the chip. Weights of LSTM and dense layers are programmed onto all 64 cores. The rows 1-4 encode the weights LSTM unit and vertical on-chip links connect their LDPUs to realize on-chip data aggregation. The LDPUs of the cores in row 4 are connected to their respective GDPU slice via on-chip links. The dense layer is mapped on rows 5-8, and on-chip vertical data aggregation is setup within rows 5-6 and rows 7-8. \textbf{c}, Measured BLEU-$n$ scores from the chip with ODP compared with software baseline and simulation results that include the hardware-measured weight noise and quantization from the PWM, LDPU, and GDPU. Only ODP was used for this network because $\Gmax$ cannot be increased beyond 80 ADC counts to stay within the linear region of the ADC, therefore TDP does not bring accuracy benefits (see Methods). The error bars represent one standard deviation over 10 inference runs. \textbf{d}, Comparison between software and chip generated captions for three exemplary images, along with the reference captions over which the BLEU scores are computed. } \label{fig:5}
\end{figure*}

The third network that we implemented is meant to assess the inference accuracy on a workload that uses all 64 cores. We chose an image caption generation task implemented with a single LSTM unit and a dense layer (4,080,384 weights and 6,080 biases), similar to the PTB LSTM network described above (Fig.~\ref{fig:5}a). Here, at the initial timestep, the features of an input image extracted by a CNN (implemented in software) are fed to the on-chip input gate. As above, the MVMs from the input and hidden gates are summed on-chip and sent to the GDPU via on-chip links. The GDPU output is sent to the dense layer that produces the first word of the caption, which is stored in an off-chip caption buffer. Then, this word is embedded, sent to the input gate, and the GDPU output is sent to the hidden gate, in order to generate the next word of the caption (Fig.~\ref{fig:5}b). The latter process is run until the entire caption is generated, for at most 24 timesteps per image (see supplementary video). The performance of the image caption generation task is evaluated using the BLEU-$n$ score~\cite{bleu}, which measures the match of word $n$-grams between a generated caption and multiple reference captions. The BLEU scores measured over the full test dataset are shown in Fig.~\ref{fig:5}c, along with three representative generated captions in Fig.~\ref{fig:5}d. The BLEU scores achieved by the chip are essentially the same, within variation, as the software ones. It is observed that captions of some images differ between software and hardware, sometimes one being better or worse than the other, as shown in the examples of Fig.~\ref{fig:5}d.  However, when averaged over all images of the dataset, the hardware BLEU scores are almost identical to the software ones.

\section{Performance}


Table \ref{exttab:1} summarizes the measured performance achieved by the chip for three use-cases: (i) the chip performing only MVMs at full utilization, (ii) the processing of one input to a deep layer of ResNet-9, and (iii) the processing of one timestep of the LSTM unit of the image captioning network (see Methods). A peak MVM throughput of 63.1 TOPS is achieved at an energy efficiency of 9.76 TOPS/W and MVM area efficiency of 1.55 TOPS/mm$^2$ for the 1-phase read mode, when all 64 cores are fully utilized to perform MVMs in parallel. For the 4-phase read mode, the former metrics are approximately 4 times lower due to the $\sim 4$x higher latency. Even then, the raw MVM throughput of the chip in 4-phase read mode is higher than the previous state-of-the-art AIMC chips based on RRAM, PCM and SRAM, and is comparable to the nor-Flash chip presented in Ref.~\onlinecite{mythic2022}, while achieving the highest accuracy on the CIFAR-10 benchmark (see Extended Data Table \ref{exttab:2}). Although the SRAM-based accelerator presented in Ref.~\onlinecite{Jia2022} shows a higher energy efficiency and throughput density, it requires the use of an off-chip weight buffer to store the total weight data of a network, leading to additional overhead during inference execution.
In contrast, our implementation enables to store a model entirely on-chip, therefore reducing external memory accesses during execution as well as being able to retain the weight data throughout power cycling. 

Finally, regarding the efficiency of the neural network layers including the digital operations and on-chip data aggregation through links, a single input to the ResNet-9 layer is processed in \unit[1.52]{$\mu$s} and consumes \unit[1.51]{$\mu$J}, whereas a single timestep of the LSTM unit is executed in \unit[1.43]{$\mu$s} while consuming \unit[5.24]{$\mu$J}. 
Reducing the precision of the digital computing from \texttt{FP16} to lower integer precision, such as \texttt{INT8}, will allow in the future to parallelize the digital computations across ADC outputs to reduce the latency gap with the MVM-only latency.

\section{Conclusions}

We have presented the IBM HERMES Project Chip, a 64-core AIMC chip based on 14-nm CMOS with backend-integrated PCM that can be used for deep neural network inference tasks. The chip can implement $> 4$M weights split across 256x256 cores, perform MVMs through AIMC, and support all non-MVM computations involved in convolutional and LSTM layers via dedicated digital units. The chip achieves a peak MVM throughput of \unit[$16.1-63.1$]{TOPS} at an energy efficiency of \unit[$2.48-9.76$]{TOPS/W} in 4-phase (high-precision) $-$ 1-phase (low-precision) read mode for 8-bit input/output MVMs. The measured MVM throughput per area of \unit[400]{GOPS/mm$^2$} in 4-phase read mode is $>15 \times$ higher than previous multi-core AIMC chips based on resistive memory, while achieving the highest CIFAR-10 accuracy of $92.81\%$ among them. Here we addressed some of the critical challenges for AIMC such as overcoming the need for application-wise re-tuning of the chip. We also demonstrated the integration of AIMC with some of the essential digital compute blocks and an on-chip communication fabric. With additional digital circuitry that enables the layer-to-layer activation transfers and intermediate activation storage in local SRAM with appropriate appending and reorganization, fully pipelined end-to-end inference workloads can be run on-chip \cite{Y2021dazziFCN}. This will be the subject of future work.
Further improvements in weight density will also be highly desirable for AIMC accelerators to be a strong competitor to existing digital solutions, which could be foreseen by integrating the PCM devices closer to the transistor-level at a denser pitch or via 3D stacking of memory layers\cite{lin2020,huo2022}.

\section*{Methods}

\subsection*{Chip and PCM device fabrication.}
Our experimental results are demonstrated on chips from \unit[300]{mm} wafers fabricated with PCM inserted in a \unit[14]{nm} back-end-of-line (BEOL) at IBM Research at Albany NanoTech. The BEOL fabrication including PCM mushroom cell formation is done on top of the foundry-sourced front-end wafers. The PCM mushroom cell is comprised of a ring heater for the bottom electrode and a doped Ge$_2$Sb$_2$Te$_5$ and top electrode film stack which is subtractively patterned to form the mushroom top. Wafers are characterized through several different in-line electrical tests to assess their quality before identifying wafers and die to send for dicing and packaging. Testing is done on discrete single structures embedded in the BEOL and also done on array diagnostic monitors (ADM) with functional peripheral circuitry. 

\subsection*{Experimental platform.}
The experimental platform is built around a single chip packaged in a 1,525 pin ball grid array (BGA).
Through controlled collapse chip connection (C4), the chip can be mounted onto an interposer printed circuit board, which is placed on a socket for testing on a dedicated test board. For every core, $72$ dedicated C4 pads provide analog and digital supply voltages in addition to a small number of pads used to implement a serial communication interface to control the core. 
The supply voltages of the cores in each row are grouped on the interposer.
The resulting partition into eight power domains reduces the required number of components on the test board while maintaining a high degree of redundancy and resilience against faulty cores, for example due to manufacturing failures.

The chip is interfaced to a hardware platform comprising one field-programmable gate array (FPGA) module and a base board. The base board provides the chip socket, chip cooling, multiple power supplies per power domain as well as the voltage and current reference sources for the chip. A Trenz TE0808-04 module with Xilinx Zynq UltraScale+ FPGA is used to implement overall system control and data management as well as the interface with the chip. The experimental platform is operated through Ethernet from a host computer, and a Python environment is used to coordinate the experiments. The base board provides current measurement for all the chip supply rails by using high-side current sensing before the voltage regulators, allowing averaged current measurements with voltmeters.

\subsection*{Weight programming.}
The gate signals of the unit-cells are connected in a diagonal fashion.  The goal of this is to allow as many devices to be programmed in parallel as possible without causing an excessive current flow on any one bit line (BL) or  source line (SL). During programming, only one gate signal is enabled, as depicted in Extended Data Fig.~\ref{extfig_prog}a. These gate selection signals are controlled by the diagonal selection module.

The programming circuit controls the application of programming current profiles to the SLs of the crossbar array. There are a total of 32 parallel digital-to-analog converters (IDACs) that can apply digitally-controlled current values to the SLs. Each IDAC is assigned to 8 consecutive SLs of either negative or positive polarity (16 in total). A programming finite-state-machine is in charge of controlling these IDACs from the configuration values written to its registers, as well as selecting the active SLs of the PCM devices to program. 

To program a set of devices, a single diagonal must be selected before the IDACs are activated. This way, current can be applied to all SLs in parallel without causing any current congestion in the chip. In practice only 32 SLs are programmed in parallel due to area and power constraints. Therefore, it takes 8 programming cycles to completely program one diagonal of devices for a particular polarity and device ID (1 or 2). Both square current pulses (RESET) and current pulses with a trailing edge (SET) are supported with configurable pulse widths and amplitudes. In the experiments presented here, we use a RESET pulse width of \unit[125]{ns} with amplitude of \unit[700]{$\mu$A}, and a SET pulse width of \unit[250]{ns} with trailing edge of \unit[50]{ns} and amplitude of \unit[125]{$\mu$A}. 

When programming a weight onto a unit-cell, all 4 PCM devices are first RESET, then the devices corresponding to the weight polarity are SET. For ODP, the SET pulses are sent only to the device 1 and for TDP to devices 1 and 2. Thereafter, the device which will receive the subsequent iterative programming pulses is selected. For ODP, device 1 is always selected. For TDP, if the target conductance value $|G_\mathrm{mn}|$ exceeds $\GSET$ of both devices 1 and 2, the device with the lowest $\GSET$ is selected for programming and the other is left as-is. Otherwise, the device with the highest $\GSET$ is selected for programming and the other is RESET. Extended Data Fig.~\ref{extfig_prog}b shows a diagram of the TDP algorithm.

Iterative programming involving a sequence of program-and-verify steps is then used to program the selected devices to the desired conductance values.\cite{Y2011papandreouISCAS} After each programming pulse, a verify step is performed, and the value of the unit-cell conductance programmed in the preceding iteration is read. The read is performed by applying long PWM pulses of \unit[$0.2$]{V} amplitude with \unit[512]{ns} width and \unit[256]{ns} precharge time to the selected diagonal, to ensure reliable read of the small individual conductance values by the ADCs. The programming current applied to the PCM device in the subsequent iteration is adapted according to the value of the error between the target level and the read value of the unit-cell conductance. The programming pulses are square pulses of \unit[125]{ns} width and amplitude varying between \unit[125]{$\mu$A} and \unit[700]{$\mu$A}. The programming sequence ends when the error between the target conductance and the programmed conductance of the unit-cell is smaller than a margin of 5 ADC counts or when the maximum number of iterations (30) has been reached.

\subsection*{ResNet-9 on CIFAR-10 training.}
The CIFAR-10 dataset~\cite{cifar} is a well-known image classification benchmark that comprises a total of 60,000 RGB images belonging to one of 10 classes. The dataset is split into a training (50,000 images) and test (10,000) set. Each image has a size of $32 \times 32$.

We use ResNet-9~\cite{resnet} since it is one of the most commonly used architectures for image classification. The architecture comprises 8 convolutional layers (all followed by a batch norm and ReLU activation layer) which can be attributed to two main blocks. Each block has two convolutional layers and a residual layer that contains another two convolutional layers. The number of filters for each convolutional layer in order of appearance is $[56,112,112,112,224,224,224,224]$. This network yields a total of 1,869,122 trainable parameters.

We first train a model without any hardware-aware training for $N=300$ epochs and later, as part of the hardware-aware training, finetune the pretrained model for additional $N=200$ epochs. Note that, both the training phases are fully executed off-chip in software. The batch size is set to $B=128$. In both phases of the training, we use SGD where the initial learning rate $\eta=0.05$ is gradually reduced using a cosine annealing scheduler with $T_\textnormal{max}=200$. In order to improve performance, we randomly crop the training images to a shape of $32 \times 32$ (padding=4) and apply Cutout~\cite{cutout} with a patch size of $4 \times 4$.
We apply the following methods during hardware-aware finetuning of the model. After every update, we clip the weights of each layer to a constant value ($\alpha=1$) around zero. The tiling of layers is simulated during training with a tile size of $256 \times 256$. In order to make the model more robust to noise, we perturb the weights of each tile according to a model simulating two times the PCM programming noise derived experimentally in Ref.~\onlinecite{nandakumarICECS}. Additionally, we add additive Gaussian noise on the output of each per-tile MVM with standard deviation of 10\% of the maximum output range. Finally, DACs and ADCs at the output of each tile are simulated during training with a resolution of 8 bits.

\subsection*{PTB-char LSTM network training.}
The Penn Treebank (PTB)~\cite{ptb} corpus is a collection of articles that is commonly used for evaluating sequence labelling approaches, as well as the language modelling capabilities of models on a per-word or per-character (PTB-char) level. We use PTB-char which has a vocabulary size of 50, i.e. there are 50 different characters in the dataset. During preprocessing, every character is embedded to a random orthogonal vector (dimension $E=128$) drawn from a standard Gaussian distribution. The training sequence of length 5,017,482 is split up into smaller sequences of length 150, which are shuffled after each epoch. The validation sequence is split up into smaller sequences of size 128, which we then use to determine the best model during training. The best model is finally evaluated on the whole test sequence of length 442,423. Since we are considering a language modelling task, the target sequence corresponds to the input sequence shifted to the left by one character.

We use a one-layer LSTM~\cite{lstm}, where the input gate projects the embedded inputs for each time-step to the LSTM cell (dimension $G=2016$). The projected input is then added to the equally shaped projection of the hidden state ($H=504$). The LSTM cell then processes the combined input and produces the next cell- and hidden state. The latter is used as input to the output layer, which applies dropout~\cite{dropout} ($p=0.25$) and maps the input to the output dimension, which is equal to the number of characters in the vocabulary. For the embedding, we use vectors sampled from a standard Gaussian that are orthogonalized using Gram-Schmidt orthogonalization. 

The training procedure is split into the standard training phase, followed by the hardware-aware finetuning. In the standard training phase, we train the model for $N=200$ epochs using the Adam~\cite{adam} optimizer with an initial learning rate of $\eta=0.01$, which is reduced by a multiplicative factor of $\gamma=0.1$ every 50 epochs. We use a batch size of $B=128$ and clip the gradients to have a maximum $l_2$-norm of 5.0 in order to avoid training instabilities. After the standard training phase, we set the dropout probability to $p=0.0$ and inject Gaussian noise into the weights during each forward pass~\cite{Joshi2020}. Given the weight matrix $W$, each weight $W_{i,j}$ follows the distribution $\mathcal{N}(W_{i,j},\zeta \cdot |W_\textnormal{max}-W_\textnormal{min}|)$, where $\zeta=0.075$ controls the amount of noise relative to the dynamic range of the weights. In order to avoid weight outliers, we clip the weights of each layer to $\alpha=2.0$ times the layers' standard deviation. We find that for randomly distributed time-steps, the inputs to the hidden gate contain performance-critical outliers that can not be clipped post-training. We therefore also clip the inputs to the hidden gate (which are also the outputs of the LSTM cell) to a constant value of $\beta = 0.06$.

\subsection*{LSTM network for image caption generation training.}
We use Flickr8k~\cite{flickr}, which is a collection of images from Flickr with 5 human-generated captions per image. The dataset is split into a training, validation, and test set with 6k, 1k, and 1k samples, respectively. Due to size limitations incurred by the hardware, we reduce the vocabulary size from the original 8,918 to 4,064 (16 cores with output size 254) by removing the least frequently used words. As a consequence, 12\% of the available captions are eliminated. Nine images are removed from the dataset altogether because every caption contains an eliminated word. Using BLEU-4, which is the most common BLEU variant, we verified that removing 12\% of the captions did not significantly alter the performance of the network ($0.1609 \rightarrow 0.1557$ for the whole and reduced dataset, respectively). As a final preprocessing step, we apply basic tokenization on the sequences.

Similar to~\cite{nic}, we extract 2,048 features from each image using an InceptionV3~\cite{inception} model that was pretrained on ImageNet~\cite{imagenet}. The extracted features are then projected to a 504-dimensional subspace, followed by batch normalization~\cite{batchnorm}. As a generative model, we use a one-layer LSTM~\cite{lstm} architecture (dimension $G=2,016$). At time $t=0$, the LSTM receives as input the embedded image features and an all-zero hidden state vector. At time $t>0$, the LSTM receives as input the embedded word (learnable embedding vectors initialized with uniform distribution) and hidden state that was generated in the previous timestep. At each timestep, the output is generated by passing the 504-dimensional hidden state of the LSTM through a dense layer, producing an output vector of shape 4,064, over which a softmax is computed. At each timestep, the word with the highest softmax probability is chosen. The LSTM runs for at most 24 timesteps or until an end-of-sequence token is generated. 

We first train a network without any hardware-aware training for $N=70$ epochs using a batch size of $B=120$. We use the Adam~\cite{adam} optimizer with an initial learning rate of $\eta = 0.001$. To avoid overfitting, we apply dropout to the activations of the input layer with a dropout probability of $p=0.5$. We then finetune the pretrained model for another 50 epochs using hardware-aware training. During this phase, we inject Gaussian noise (the same type as for PTB-Char, $\zeta=0.08$) and evaluate with slightly less noise ($\zeta = 0.035$). After every update, we clip the weights to $\alpha = 2.0$ standard deviations and apply $l_2$-regularization with $\lambda=1\mathrm{e}{-4}$. To avoid exploding gradients, we clip the gradient $l_2$-norm to 0.5. During training, we maximize the sum of the negative log likelihood of the correct word at each step. As evaluation metric, we use the BLEU~\cite{bleu} score, which measures the precision of word $n$-grams between a generated and multiple reference captions. We report the $n$-gram BLEU score for $n \in \{1,2,3,4\}$.

\subsection*{Weight mapping onto cores and functional modeling.}

The mapping of weights onto the chip is done by converting a weight matrix for each layer into 256x256 conductance sub-matrices, one for each core the layer is mapped onto. The weight matrix is split into the smallest possible amount of sub-matrices which have the same number of rows and columns, both less than 256. For example, the 2016x224 weight matrix of a deep layer of ResNet-9 is converted into 8 252x224 sub-matrices. Each sub-matrix is then reshaped into a 256x256 matrix by zero-filling. Finally, the weight values of each sub-matrix are converted to conductance values using Eq.~\eqref{eq:weight_to_G}, with $\Gmax$ being determined on a per-core basis. The $\Gmax$ value of a core is set to the minimum value between the maximum conductance supported by the unit-cell (80 ADC counts for ODP and 160 ADC counts for TDP, as determined in Fig.~\ref{fig:2}a) and the maximum conductance that leads to currents of all BLs to be less than the maximum current supported by the ADC (approximately \unit[100]{$\mu$A}). For higher BL currents, the ADC response starts to saturate, which introduces nonlinearity in the resulting MVMs and therefore must be avoided. This additional constraint means that for some layers the full unit-cell conductance range cannot be utilized. For ODP, the $\Gmax$ value used for all networks is 80 except for the image captioning network, where the hidden gate layer uses 70 and the output layer values between 40 and 80 depending on the core. For ResNet-9 with TDP, we use $\Gmax$ of 160 for conv0 and dense layers, values between 80 and 120 for conv1-4 and values between 120 and 140 for conv5-7. For PTB-char with TDP, we use $\Gmax$ of 100 for the LSTM unit and 120 for the output layer. Therefore, only ResNet-9 takes significant advantage of TDP, because the lower utilization of the cores (especially for conv0 and dense layers) allows to use larger $\Gmax$ values that improve the MVM accuracy. 

Given the conductance mapping provided by the above procedure, we built a functional model of the chip that includes the quantization from the digital peripheral circuits and the hardware-measured weight noise. The quantization model includes the quantization resulting from all data conversions performed by the chip during inference execution. It includes the 8-bit input quantization from PWM, 12-bit quantization from ADC, 8-bit data conversion from LDPU and a GDPU functional model (including the discrete tanh LUTs and 8-bit output data conversion). The weight noise model includes the per-core weight error measured when performing MVMs with the network weights being programmed. We perform multiple MVMs on each core and compute the resulting weight error $\mathrm{std}(W-\hat{W})/\Wmax$, where $\mathrm{std}(W)$ is the standard deviation computed over all elements of $W$, as shown in Fig.~\ref{fig:2}c. We then fit the experimental weight error data points as a function of target weight with a polynomial function for each core. Weight noise during inference is simulated by injecting Gaussian noise to the synaptic weights with a weight-dependent standard deviation given by this polynomial function. 

\subsection*{Power measurements.}

We measured the chip energy efficiency and latency for three different use-cases. The first one is the chip performing only MVMs at full utilization (all 64 cores active in parallel), which provides the maximal MVM efficiency of the chip that can be compared with other works. The reported MVM energy includes the PCM crossbar array, ADCs, and PWMs (operations performed by the LDPU and GDPU used in neural network executions, such as activation functions, are not included in the MVM energy). The second use-case is the processing of one input to a deep layer of ResNet-9 that spans 8 cores, including the MVMs, data aggregation with on-chip links and LDPU operations. The third use-case is the processing of one timestep of the LSTM unit of the image caption generation network, including the MVMs, data aggregation with on-chip links, LDPU and GDPU operations. 

For each use-case, the power consumed by the chip is measured both in standby mode (static) and while performing the computations (dynamic) to calculate the energy. The average current drawn by the ADC and digital supplies operating at \unit[0.85]{V} is measured using voltmeters (Fluke models 87V and 185). The dynamic power consumption is obtained for all the monitored supplies by sending a stream of MVM commands with a period of \unit[5]{$\mu$s}, and measuring the averaged power during continuous operation of a few seconds. The dynamic energy is computed by multiplying the dynamic power by the MVM command period. The total energy is calculated by adding the static and dynamic contributions for all monitored supplies. The latency is computed by running the register transfer level (RTL) code for each use-case. 

\newpage
\section*{References}
\bibliography{papers_mem}

\section*{Acknowledgments}
We thank Geoffrey W. Burr, Markus B\"{u}hler, Thilo Maurer, Antje M\"{u}ller, Yasuteru Kohda, Kohji Hosakawa, Fee Li Lie, Feng Liu, Ted Levin and Tarl Gordon for assistance with the chip design; Atsuya Okazaki, Hirayuki Mori and Marc Bergendahl for assistance with the chip packaging; Jordi Fornt Mas, Giorgio Cristiano and Joachim Paret for chip testing and simulation; Fr\'{e}d\'{e}ric Odermatt, Irem Boybat, S. R. Nandakumar, Christophe Piveteau, Corey Lammie, Hadjer Benmeziane and Malte Rasch for help with network deployment on the chip; Angeliki Pantazi, Robert Haas, Alessandro Curioni, Sidney Tsai, Wilfried Haensch, Jeff Burns, Rama Divakaruni and Mukesh Khare for managerial support. We would also like to thank Prof. Luca Benini and Prof. Bipin Rajendran for their support with supervising students. This work was supported by the IBM Research AI Hardware Center. We also acknowledge partial funding from the European Research Council (ERC) under the European Union’s Horizon 2020 research and innovation programme (grant agreement numbers 682675 and 966764).


\section*{Competing interests}
The authors declare no competing interests.

\clearpage
\clearpage

\clearpage

\clearpage

\clearpage

\clearpage

\begin{table}[t!]
\centering
\begin{tabular}{l r | l r | l r | l}
 \toprule
  & \multicolumn{2}{c}{MVM at max. utilization} & \multicolumn{2}{c}{ResNet-9 layer} & \multicolumn{2}{c}{One timestep of LSTM unit} \\ 
 \midrule
 Number of cores used & \multicolumn{2}{c}{64} & \multicolumn{2}{c}{8} & \multicolumn{2}{c}{32} \\
 Average core utilization &  \multicolumn{2}{c}{100\%} & \multicolumn{2}{c}{86\%} & \multicolumn{2}{c}{97\%} \\
 \midrule
 Read mode\footnotemark[1] & 1-phase & 4-phase & 1-phase & 4-phase & 1-phase & 4-phase \\
 \midrule
 Peak MVM throughput (TOPS)\footnotemark[2] & 63.1 & 16.1 & 6.79 & 1.74 & 30.6 & 7.82 \\ 
 Peak MVM energy efficiency (TOPS/W)\footnotemark[2] & 9.76 & 2.48 & 6.88 & 1.74 & 9.34 & 2.37 \\ 
 MVM area efficiency (TOPS/mm$^2$)\footnotemark[2] & 1.55 & 0.40 & 1.34 & 0.34 & 1.50 & 0.38 \\ 
 Energy ($\mu$J) & 0.86 & 3.38 & 0.97 & 1.51 & 3.46 & 5.24 \\ 
 Latency (ns) & 133 & 520 & 1131 & 1518 & 1043 & 1430 \\
 \bottomrule
 \footnotetext{All MVM and inference accuracies are reported for 4-phase mode. }
 \footnotetext{Throughput, energy, and area consider only the MVM operations performed by the cores. 1 multiply-and-accumulate = 2 OPs. }
\end{tabular}
\caption{Measured speed and energy efficiency of IBM HERMES Project Chip. } \label{tab:1}
\end{table}

\clearpage
\renewcommand{\figurename}{EXTENDED DATA FIG.}
\setcounter{figure}{0}

\begin{figure*}[t!]
\centering
\begin{tabular}{c}
\includegraphics[width = 0.99\columnwidth]{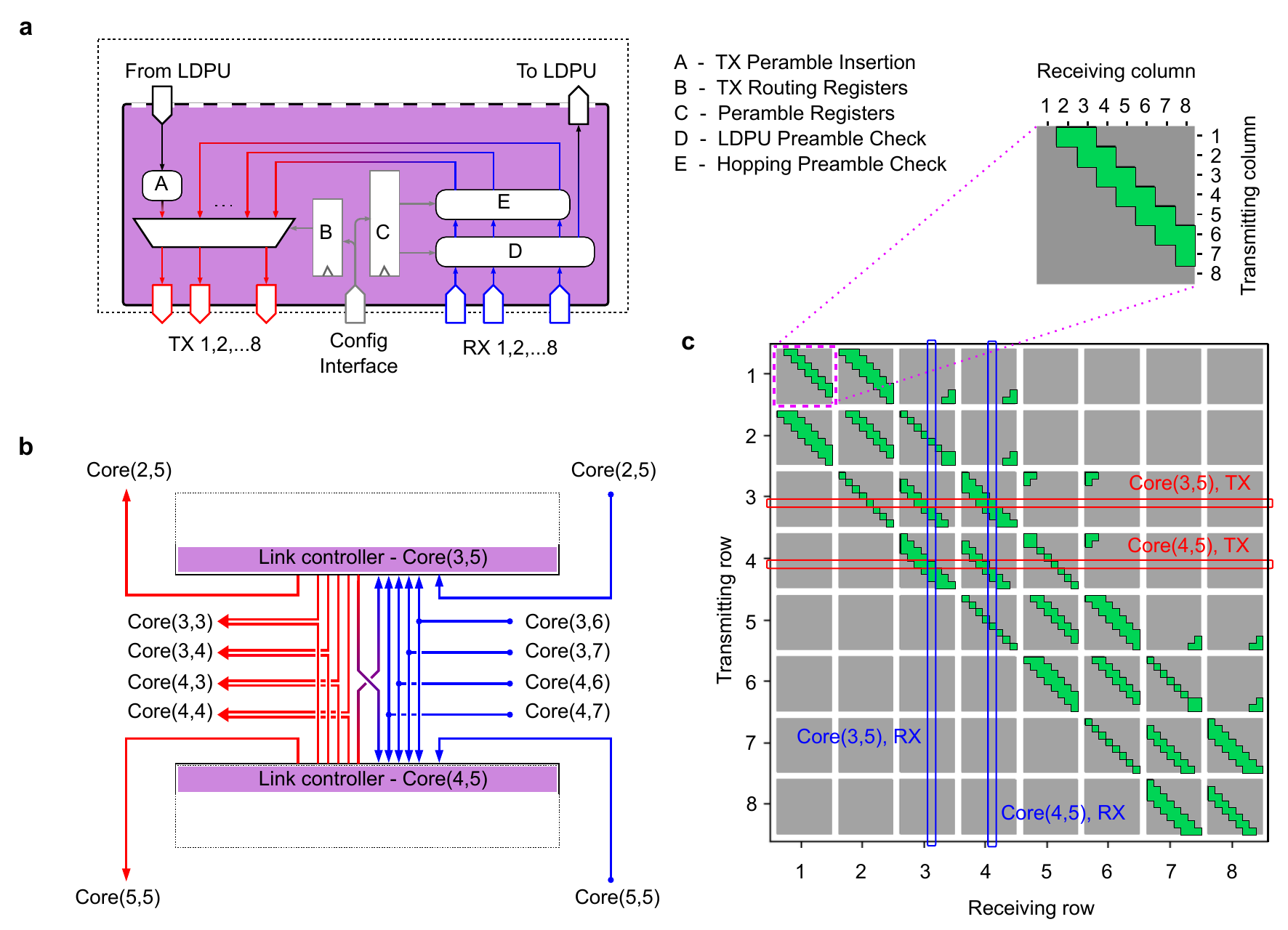}
\end{tabular}
\caption{\textbf{Digital communication fabric. } \textbf{a}, Schematic of link controller. The dotted bounding box refers to the core boundary. \textbf{b}, Possible link connections for Core(3,5) and Core(4,5), where the notation Core($r$,$c$) refers to the core located at row $r$ and column $c$ in Fig.~\ref{fig:1}b. \textbf{c}, Link connections for the entire chip (available connections are denoted in green color). The RX and TX connections for Core(3,5) and Core(4,5) shown in \textbf{b} are indicated. }  \label{extfig_link}
\end{figure*}
\clearpage

\begin{figure*}[t!]
\centering
\begin{tabular}{c}
\includegraphics[width = 0.99\columnwidth]{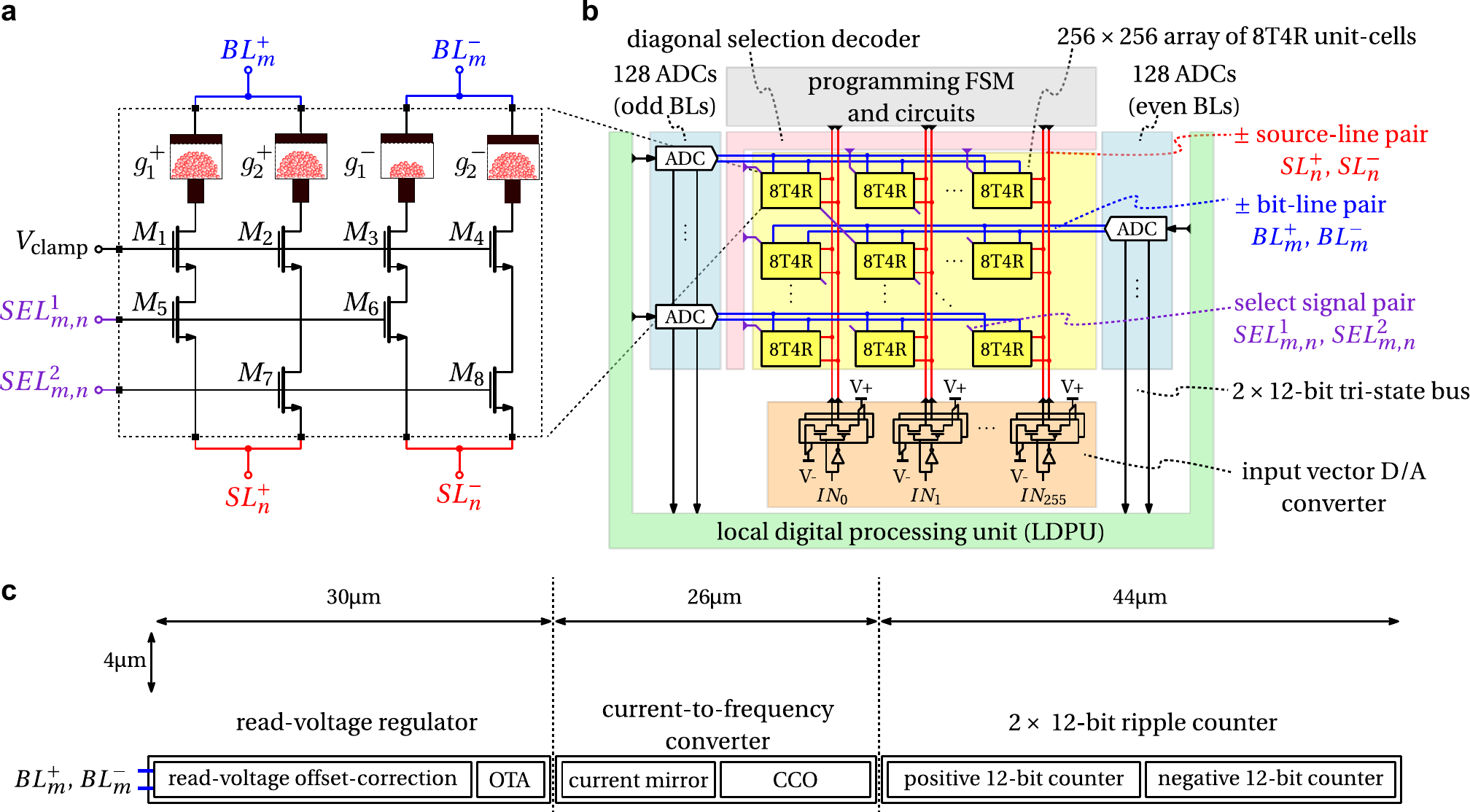}
\end{tabular}
\caption{
	\textbf{PCM crossbar array. }
	\textbf{a}, Schematic of 8T4R unit-cell. The top electrodes of the conductance pairs of each polarity connect to separate bit lines $BL_{m}^{+}$, $BL_{m}^{-}$ and the sources of their lower access-transistors connect to separate source lines $SL_{n}^{+}$, $SL_{n}^{-}$. Thus, the devices in a conductance pair are weighted with equal significance and the total conductance per unit-cell becomes: $\left(g_{1}^{+}+g_{2}^{+}\right)-\left(g_{1}^{-}+g_{2}^{-}\right)$.
	\textbf{b}, Schematic of PCM crossbar array. To program the PCM devices, the dedicated per-core programming FSM instructs the diagonal selection decoder to enable one diagonal of cells that contains the devices that are to be programmed. The diagonal selection decoder controls the ${SEL}_{m,n}^{1}$ and ${SEL}_{m,n}^{2}$ signals in the unit-cell, which are routed diagonally throughout the array. The selected devices are programmed by the current-steering DAC-based programming units located on top of the PCM array. To perform an MVM, the 256 inputs to the crossbar array ($IN_0-IN_{255}$) are applied via the red source lines (SLs) to the 8T4R cells. The resulting bit line (BL) currents are summed up on the blue wires and read by the ADCs that flank the crossbar array on the left and right.
	\textbf{c}, Layout of one ADC. The block diagram that is shown below the layout illustrates the various components of the ADC, namely, the read voltage regulator, the current-to-frequency converter, and the $2\times{}$12-bit ripple counter.
}\label{extfig_core}
\end{figure*}
\clearpage

\begin{figure*}[t!]
\centering
\begin{tabular}{c}
\includegraphics[width = 0.99\columnwidth]{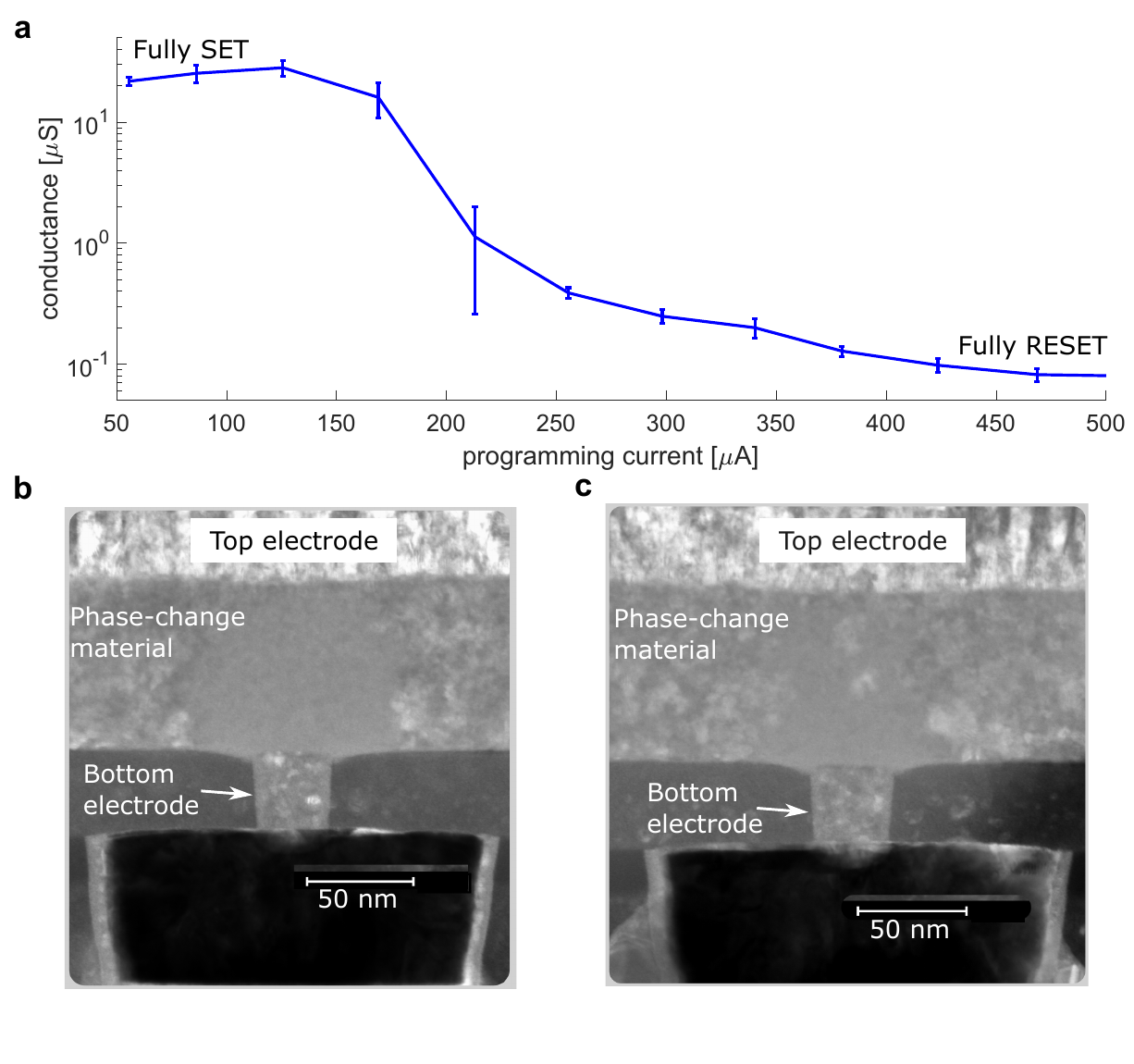}
\end{tabular}
\caption{
	\textbf{PCM device.}
	\textbf{a}, A typical programming curve indicating the programmed device conductance as a function of the programming current. The measurements are repeated 10 times on a single device and the error bar indicates the standard deviation. The device conductance is determined by the phase configuration within the PCM device and in particular, the size of the amorphous region.
	\textbf{b}, Low-angle annular darkfield (LAADF) scanning transmission electron microscope (STEM) image of a fully \textsc{reset} PCM device showing a substantially large amorphous region that fully blocks the bottom electrode. LAADF enables the imaging of the amorphous region with high resolution \cite{Y2021liISTFA}. 
	\textbf{c}, LAADF of a partially \textsc{reset} PCM device showing a much smaller amorphous region. The synaptic weights are stored in an analog manner in terms of these phase configurations and the resulting conductance values.
}\label{extfig_device}
\end{figure*}
\clearpage

\begin{figure*}[t!]
\centering
\begin{tabular}{c}
\includegraphics[width = 0.99\columnwidth]{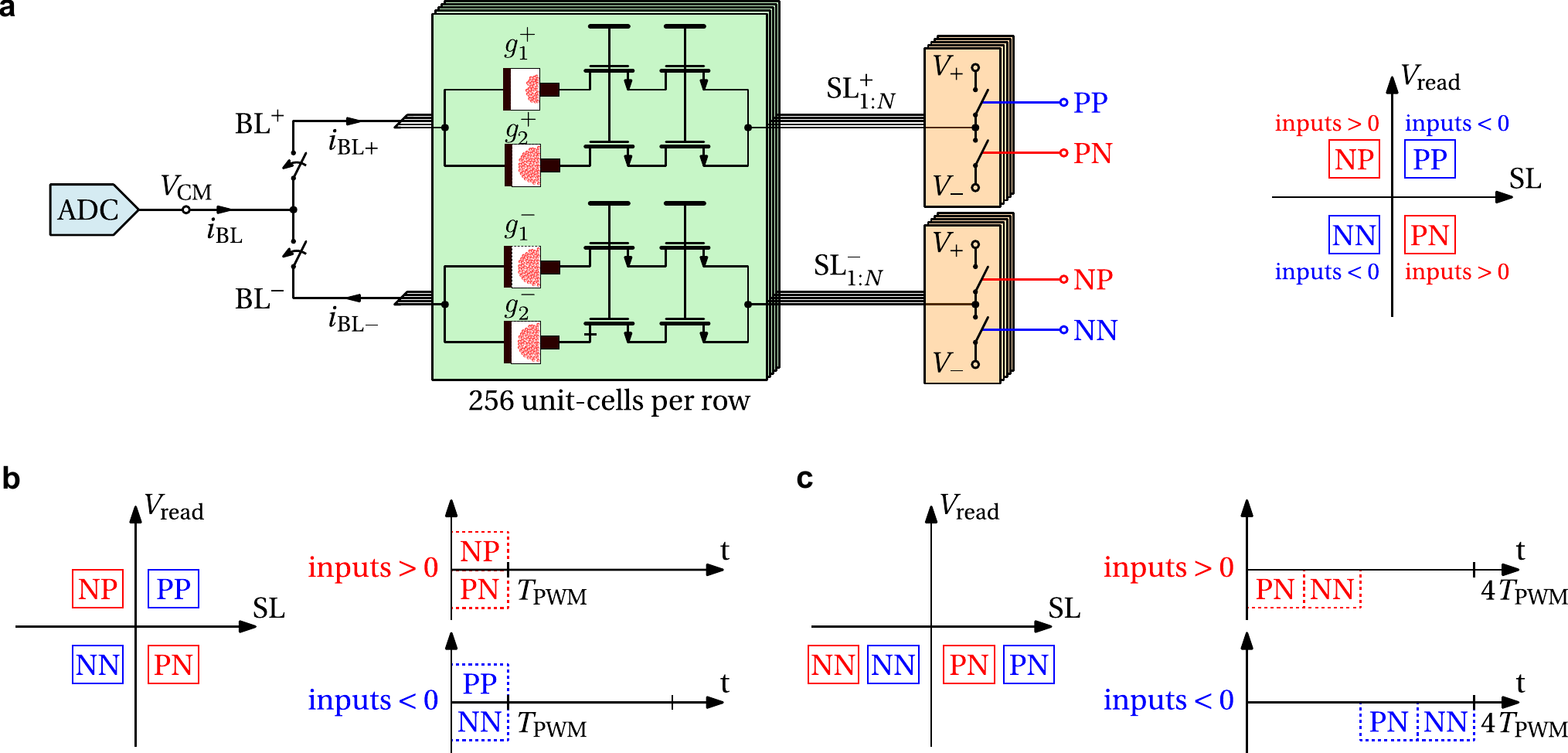}
\end{tabular}
\caption{%
    \textbf{Input modulation modes for MVM. }
    \textbf{a}, Full array read procedure for MVMs showing the connection between ADC, unit-cells, and input modulator switches.
    Signals PP and PN connect the positive source lines $SL_{1:N}^{+}$ to the positive potential $V^+$ and negative potential $V^{-}$, respectively. For NP and NN it is vice versa.
    \textbf{b}, 1-phase modulation mode. Inputs of positive and negative polarity are applied to weights of positive and negative polarity in one modulation cycle $T_\mathrm{PWM}$.
    \textbf{c}, 4-phase modulation mode. Inputs of positive and negative polarity are applied individually to weights of positive and negative polarity in four modulation cycles. }\label{extfig_read}
\end{figure*}
\clearpage

\begin{figure*}[t!]
\centering
\begin{tabular}{c}
\includegraphics[width = 0.99\columnwidth]{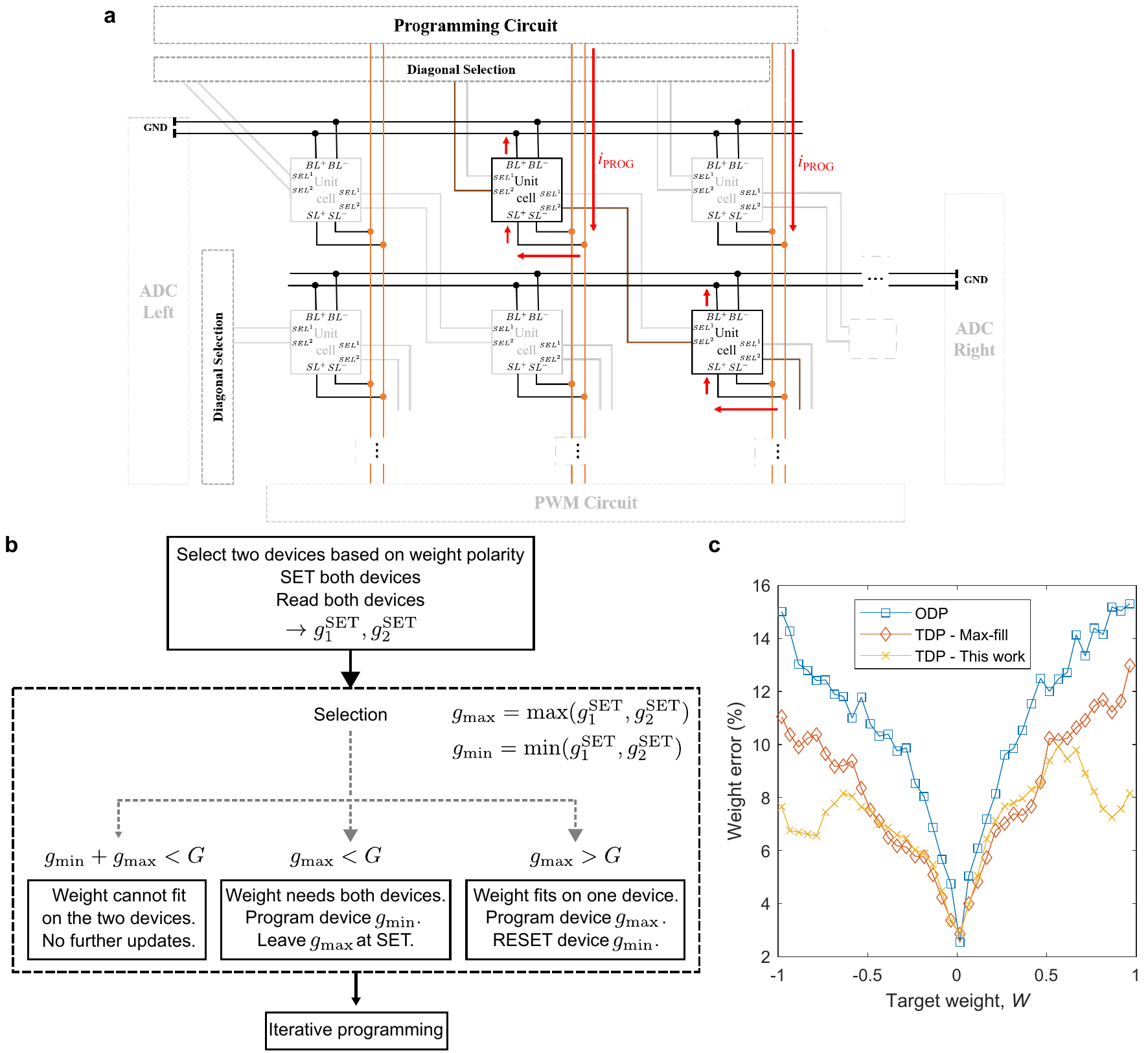}
\end{tabular}
\caption{\textbf{Weight programming procedure. } \textbf{a}, Crossbar array during programming. \textbf{b}, Proposed TDP algorithm to program a target conductance value $G$ on a unit-cell. \textbf{c}, Weight error comparison between TDP of this work and previous approaches. TDP - Max-fill refers to programming the two devices with iterative programming up to the ODP $\Gmax$, as proposed in Ref.~\onlinecite{legallo2022}. Due to the wide SET distribution shown in Fig.~\ref{fig:2}a, some devices in the core either cannot achieve $\Gmax$, or conversely could be programmed to much higher conductance values than $\Gmax$. Therefore, the latter approach leads to programming inaccuracies resulting from either under-utilizing the conductance range of individual devices or from devices that cannot reach $\Gmax$. The proposed TDP algorithm solves this issue by using the readout SET conductance of the devices of the unit-cell to map the weight. }  \label{extfig_prog}
\end{figure*}

\clearpage
\renewcommand{\tablename}{EXTENDED DATA TABLE}
\setcounter{table}{0}

\begin{table}
\centering
$\begin{array}{l|c}
	\toprule
	\text{Name}                                        & \text{Value}                                             \\ 
\midrule
	\text{Technology}                                  & \text{14-nm CMOS + IBM PCM}                   \\
	\text{Supply voltage}                              & 0.85 - \unit[0.95]{V}  
	        \\
	\text{Chip Area}                                   & \unit[12]{mm} \times \unit[12]{mm} = \unit[144]{mm^2}    \\
	\text{Single core area}                            & \unit[1.2]{mm}\times\unit[1.16]{mm} = \unit[1.392]{mm^2} \\
	\text{MVM area per core}                           & \unit[0.870]{mm}\times\unit[0.730]{mm} = \unit[0.635]{mm^2}   \\       
	\text{GDPU macro area}                             & \unit[0.809]{mm}\times\unit[0.347]{mm} = \unit[0.281]{mm^2}  \\      
	\text{Number of cores}                             & 64                                                       \\
	\text{Number of GDPUs}                             & 8                                                        \\ 
\midrule
	\text{Number of C4 pads per core}                  & 72                                                       \\
	\text{Total number of C4 pads}                     & 5,616                                                    \\
	\text{External contact pads}                       & 1,520                                                    \\ 
\midrule
	\text{Maximum MVM clock frequency}                 & \unit[1]{GHz}     
	             \\
	\text{Maximum LDPU/GDPU clock frequency}           & \unit[400]{MHz}   
	             \\
	\text{Maximum link data-rate}                      & \unit[3.2]{Gbit/s}                                          \\ 
\midrule
	\text{Number of unit-cells per core}               & 256\times{}256 = 65,536                                  \\
	\text{Number of PCM devices per unit cell}         & 4                                                        \\
	\text{Number of PCM devices per core}              & 256\times{}256\times{}4 = 262,144                        \\
	\text{Number of PCM devices per chip}              & 256\times{}256\times{}4\times{}64 = 16,777,216           \\
	\text{Nominal PCM resistance range per PCM device} & \unit[100]{k\Omega}-\unit[10]{M\Omega}                   \\ 
	\text{Resolution of the time-encoded inputs}       & \unit[8]{bit}~(\unit[1]{ns}/\text{LSB})    
	          \\
	\text{Read voltage range}                          & 100-\unit[400]{mV}                                         \\ 
	\text{Number of {ADC}s per core}                   & \text{256  (1 per row)}                                    \\
	\text{ADC output bits}                             & 2\times\unit[12]{bit}                                    \\
	\text{Number of write heads per core}              & 32                                                       \\
	\text{Max. supported write current per write-head} & \unit[800]{\mu{}A}                                       \\  
	\bottomrule
\end{array}$
\caption{Summary of IBM HERMES Project Chip specifications. } \label{exttab:1}
\end{table}

\clearpage
\begin{table}
\footnotesize
\centering
\begin{tabular}{ l|c | c c c c c c }
 \toprule
  & \multicolumn{2}{c}{This work} & ISSCC '22 [\onlinecite{Khwa2022}]& Nature '22 [\onlinecite{wan2022compute}] & Nat. Elec. '21 [\onlinecite{hung2021four}] & ISSCC '22 [\onlinecite{mythic2022}] & JSSC '22 [\onlinecite{Jia2022}]\\
  & 1-phase read & 4-phase read & & & & \\ 
 \midrule
 CMOS technology & \multicolumn{2}{c}{14 nm} & 40nm & 130 nm & 22 nm & 40 nm & 16nm \\ 
 AIMC device  & \multicolumn{2}{c}{PCM} & PCM & RRAM & RRAM & nor-Flash & SRAM\\ 
 Chip area (mm$^2$) & \multicolumn{2}{c}{144} & 18 & 159 & 6 & 190 & 25\\ 
 Number of cores (core size) & \multicolumn{2}{c}{64 (256x256)} & 8 (256x1024) & 48 (256x256) & 8 (1024x512) & 76 (1024x1024) & 16 (1152x256)\\  
 Number of weights & \multicolumn{2}{c}{4.2M} & 400k & 1.6M & 524k & 80M & 1.2M\\ 
 Input/weight/output precision & \multicolumn{2}{c}{8b/Analog/8b} & 8b/8b/19b & 4b/Analog/6b & 8b/8b/14b & 8b/Analog/8b & 4b/4b/8b\\
 CIFAR-10 accuracy & - & 92.81\% & 91.89\% & 85.66\% & 92.01\% & - & 91.51\% \\
 Peak MVM throughput (TOPS) & 63.1 & 16.1 & 0.475 & 0.754 & 0.0337 & 16.6 & 11.8 \\
 MVM TOPS/W & 9.76 & 2.48 & 20.5 & 16 & 15.6 & 5.2 & 121 \\
 MVM GOPS/mm$^2$ & 1550 & 400 & 26.4 & 4.7 & 5.61 & - & 2670 \\
 \midrule
 Non-MVM operations & \multicolumn{2}{c}{ReLU/sigmoid/tanh, batch} &  &  &  & All (1 RISC-V & All CNN/LSTM\\
 supported & \multicolumn{2}{c}{norm., LSTM hidden state,} & - & ReLU/sigmoid/tanh & - & processor & operations (SIMD \\
  & \multicolumn{2}{c}{partial sum accumulation} &  &  &  &  per core) & ALU + LUTs) \\
 \bottomrule
\end{tabular}
\caption{Comparison of IBM HERMES Project Chip with other multi-core AIMC chips demonstrating neural network inference. } \label{exttab:2}
\end{table}

\end{document}